\documentclass[preprint,aps,tightenlines,floatfix,showpacs]{revtex4}
\usepackage{graphics}
\usepackage{bm}

\begin{document}
\title{Probing the densities of the Sn isotopes}
\author{K. Amos}
\email{amos@physics.unimelb.edu.au}
\affiliation{School of Physics, University of Melbourne, 
Victoria 3010, Australia}
\author{S. Karataglidis}
\email{kara@physics.unimelb.edu.au}
\affiliation{School of Physics, University of Melbourne, 
Victoria 3010, Australia}
\author{J. Dobaczewski}
\email{Jacek.Dobaczewski@fuw.edu.pl}
\affiliation{Institute of Theoretical Physics, Warsaw University,
PL-00-681, Warsaw, Poland}
\date{\today}

\begin{abstract}
Proton and neutron densities have been obtained for the even-even
isotopes of Sn from $^{100}$Sn to $^{176}$Sn using a
Hartree-Fock-Bogoliubov model with a Skyrme interaction. The matter
densities so defined have been used with realistic nucleon-nucleon
interactions in a folding model to specify optical potentials for the
elastic scattering of protons with energies in the range 40 to 200
MeV. Those potentials have been used to make predictions of the
differential cross sections and spin observables for proton
scattering. As the target mass increases, the emergence of the neutron
skin in the Sn isotopes is revealed by marked effects in the
differential cross section. Comparisons with available data show how
similar scattering data for the neutron-rich isotopes may provide
constraints for the model structures.
\end{abstract}
\maketitle

\section{introduction}

One of the goals of modern nuclear physics is to understand the
structures of nuclei far from the line of stability and at extremes of
isospin. Thus far such studies have largely dealt with the light
nuclei for which radioactive beams have been available at facilities
such as TRIUMF/ISAC, NSCL, RIKEN, and CERN/ISOLDE. Early experiments
have led to the identification of novel structures of which the halo
\cite{Ha95} is the best known after its identification in $^{11}$Li.
In light of projected experimental facilities using radioactive ion
beams to extend such investigations to heavy nuclei, there have been
advances in the theoretical study of neutron rich nuclei across the
whole mass range (see refs.~\cite{Ha95,Ro92,Mu93,Ri94,Do98} for
reviews). Studies of such nuclei have topical interest since nuclei
far from stability play an important role in nucleo-synthesis with
short-lived species formed as part of the $r$- and $rp$- processes.
Notably the structure of such nuclei determine the rates at which
$\alpha$- and nucleon-capture reactions proceed against the interplay
of weak decays and photo-disintegration; key elements in the
determination of the abundances of nuclei in the universe.

Of course, these exotic systems are of interest in their own right
given that they may exhibit forms of nuclear matter quite different
from those of stable isotopes. Of particular interest are the separate
density distributions of proton and neutrons. As well as halos,
heavier systems may lead to the identification of pronounced neutron
skins \cite{Mi00} and/or to dramatic changes in nuclear shell
structure \cite{Do94}.

An important quantity by which the model structures are tested is the
ground state density. For stable nuclei, one usually seeks information
about that from the elastic scattering of electrons. The electron
scattering form factors so determined are measures of the charge and
current densities of the nucleus. Complementing that information,
which primarily focuses on the proton density itself, are analyses of
data from the elastic scattering of nucleons. Nucleon scattering
probes the matter density of the nucleus and, at the energies we
consider of the neutron matter distribution in particular. That is so
since the effective nucleon-nucleon ($NN$) interaction is strongest in
the isoscalar $^3S_1$ channel~\cite{Am00}. Of course there are
non-negligible contributions to proton scattering from their
interactions with the bound protons and such are included in all
calculations we have made. But as evident from the recent
study~\cite{Ka02} of proton and neutron elastic scattering at the same
energies from ${}^{208}$Pb, the differential cross sections from
proton (neutron) scattering do reflect changes primarily made in the
distributions of neutron (proton) matter distributions of the target.
Indeed they did so sufficiently well that cross section data could be
used to estimate the neutron skin thickness of $^{208}$Pb. For
radioactive nuclei, the only available equivalent measure of the
ground state density comes from the scattering of nuclei from hydrogen
which, in the inverse kinematics, corresponds to proton scattering
from the nucleus as target.

For nuclei above the $fp$-shell, mean-field models of structure are at
the forefront of current studies of the ground state densities.
Usually those calculations are made in a relativistic Hartree or
Hartree-Fock model \cite{Se86,Ce86,Ni92,Ri96,Bu02} or a non-relativistic
Skyrme-Hartree-Fock-Bogoliubov model \cite{Ri80,Br00,Do01,Be03}. This
mean-field approach works as the ground state properties for heavy
nuclei arise more generally from the bulk properties of the density as
opposed to single particle (SP) properties. As one approaches the drip
lines, however, surface properties become more important and, in
concert, so do SP densities.

As noted above, a test case has been to use nucleon elastic scattering
to determine/extract the neutron skin thickness in $^{208}$Pb.
Karataglidis \textit{et al.}~\cite{Ka02} have shown that the
differential cross sections from the elastic scattering of 200~MeV
protons from $^{208}$Pb suggests a neutron skin thickness in
$^{208}$Pb of $\sim 0.17$~fm; a result that has been confirmed
recently~\cite{Cl03}. More importantly, the study by Karataglidis
\textit{et al.}~\cite{Ka02} also showed that such scattering analyses
could select between disparate model structures that have the same
root mean square (rms) radii. By itself, the rms radius is not an
adequate indicator of the validity of a model structure. To use the
scattering data to differentiate between models of structure, a
predictive theory of nucleon-nucleus ($NA$) scattering was needed, and
such has been developed in the past decade~\cite{Am00}. That theory is
\textit{direct} in that all quantities required are defined \textit{a
priori} with no \textit{a posteriori} adjustment of results. With the
nucleus viewed as a system of $A$ nucleons, $NA$ scattering is
determined by an optical potential formed by a folding process. Such
microscopic approaches defining the $NA$ optical potential have been
quite successful in predicting both angle-dependent and integral
observables of elastic scattering~\cite{Am00,De01}. It is important to
note that distinction in scattering cross sections resulting from use
of different ``sensible" models of structure, are in the details and
particularly it seems with results in the region of 1 to 2.5 fm$^{-1}$
momentum transfer values. Thus one needs as complete a calculation of
the optical potentials as possible and to use it with no further
approximation if possible. The contribution to scattering of the
knock-out exchange amplitudes for all energies are such that it is not
wise to use an equivalent localization of the associated nonlocal
terms. We do not and so solve the nonlocal Schr\"odinger equations
directly. However, that also means that we must have the single
particle bound state wave functions (or equivalently the complete one
body density matrix elements) and not just the matter densities from
structure.

The Sn isotopes are of interest for current structure studies. First
many of the set that span the quite extensive range of mass between
the nucleon drip lines can be formed as radioactive ions with some in
numbers sufficient to perform scattering experiments. Measurements of
proton scattering from ${}^{114,116,118,120,122,124}$Sn have been
reported for many energies and most recently at an energy of $295A$
MeV~\cite{Me02,Te03}. Using a relativistic impulse approximation with
model (sum of Gaussians) matter densities, analyses~\cite{Te03} of the
295 MeV (preliminary) data indicated~\cite{Te03} that the isotopes
from mass 120 on had a noticeable effect from neutron occupancies of
the $3s_{\frac{1}{2}}$ shell. Their analyses also indicated that
scattering data revealed a gradual change in neutron densities as the
mass of the isotope increased. Matching the data needed inclusion of
nuclear medium effects on the ($NN$) coupling constants and exchanged
meson masses altering them from those of free $NN$ scattering. Not
only are those data preliminary, but the energy is on the high side of
the range for which we are confident that our current
non-relativistic, $g$-folding method of defining optical potentials
microscopically is valid. Notably we believe that the method by which
we define effective $NN$ interactions at and above 295 MeV may need
correction. Thus we have made calculations of scattering of protons
from many Sn isotopes at lower energies using the program
DWBA98~\cite{Ra98} that forms $g$-folding optical
potentials~\cite{Am00}. The same program then solves the associated
nonlocal Schr\"odinger equations. Specifically we have calculated
cross sections for the scattering of all the even mass isotopes $^{100
- 176}$Sn from hydrogen at an energy of $200A$~MeV. The relevant
optical potentials were formed by folding realistic effective $NN$
interactions with the details of densities of those nuclei given by
the mean-field models of their structure. With optical potentials so
formed previously, differential cross sections and spin observables
for proton scattering (and at 200~MeV specifically) from diverse
targets ranging from $^3$He to $^{238}$U have been predicted and found
to agree very well with data; providing that the structure with which
the effective $NN$ interactions are folded is appropriate. The $NN$
effective interactions were determined from $NN$ $g$ matrices
[solutions of Bruckner-Bethe-Goldstone (BBG) equations for nuclear
matter] found from realistic free $NN$ forces. Details of the
specifications of those effective $NN$ interactions, of the folding
process that gives the optical potential, and of the successful
predictions of differential cross sections and analyzing powers from
the scattering of protons at diverse energies and from diverse mass
targets, have been summarized before~\cite{Am00} and so are not
repeated herein. However of note is that with confidence in the chosen
effective $NN$ interactions and in the applicability of the
$g$-folding method, the evaluations when compared with data serve as a
test of the putative model structures. Hence a prime purpose of these
studies has been to note characteristics in scattering predictions
linked to isotope change in the Sn nuclei if they are described
appropriately by the Hartree-Fock-Bogoliubov model with a Skyrme
interaction. Of course, only access to appropriate data will allow any
critique on the specific mean-field structures of the Sn isotopes.

In the next section, we outline the structure models used in our
scattering calculations and present details of the resulting proton
and neutron matter distributions. In Section~\ref{res-scat} we show
the results found using those structures and an established effective
$NN$ force in nuclear matter in generating $g$-folding optical
potentials for a wide range of Sn isotopes scattering from hydrogen.
An energy of $200A$~MeV has been used. In Section~\ref{res-data}
application of structure and scattering models is made for a number of
cases for which proton elastic scattering data are available.
Conclusions are presented in Section~\ref{conclude}.

\section{Models of structure of the Sn isotopes}
\label{models}

In the present study, the properties of even-even Sn isotopes (masses
100 to 176) are described using the spherical mean-field
Hartree-Fock-Bruckner model~\cite{Ri80}. Two Skyrme interactions have
been used. As details of the method have been given in detail
elsewhere~\cite{Do84,Do96,Do01}, only features pertaining to the
particular calculations made and used herein are given. We have used
two parameterizations of the Skyrme force, SkP~\cite{Do84} and
SLy4~\cite{Ch97}, which, in the past, have given appropriate
descriptions of bulk nuclear properties. They differ by the input
values of the nuclear-matter effective mass, being $m^*/m = 1$ and 0.7
respectively. The zero-range density-dependent pairing force was used
in the particle-particle channel and with the form that is
intermediate between volume and surface attraction~\cite{Do01}. A
large positive energy phase space of 60~MeV was taken for which the
pairing-force strengths of $V_0=-286.20$ and $-212.94$~MeV~fm$^{-3}$
were obtained in the SkP and SLy4 cases respectively. Those strengths
result from using a standard adjustment~\cite{Do95} of the neutron
pairing gap in $^{120}$Sn.

Both model structures give good agreement with the two-neutron
separation energies $S_{2N}$ and neutron pairing gaps $\Delta_N$
extracted from data from even mass Sn isotopes having neutron numbers
between the magic numbers $N=50$ and $N=82$. At the magic numbers, the
calculated gaps vanish due to the known effect of an unphysical, too
sudden, pairing phase transition. The size of the jump of $S_{2N}$ at
$N=82$ is slightly underestimated (overestimated) by the SkP (SLy4)
model. That is an effect of the different values of the effective
mass. Nevertheless, the overall quality of agreement between theory
and experiment suffices for us to consider the two models to be useful
for extrapolations to describe unknown heavy Sn isotopes. Within the
two models, the heaviest two-neutron-bound isotope is predicted to be
mass 172(174) for SkP(Sly4).

Spatial properties of neutron and proton density distributions are of
special interest in a number of contexts. In the form of one body
density matrix elements (OBDME) and the nucleon SP wave functions,
they are central aspects of a predictive method of calculating their
scattering from Hydrogen targets. Of course, geometric aspects of
nuclei, and of the Sn isotopes in particular, have been found in
various ways the past~\cite{Mi00}. Often they have been defined by
using the Helm model and the locations of maxima and minima of
associated form factors. Also bulk density values sometimes have
assessed from averaging density functions from the structure models.
Therefore care need be exercised in equating radii and other bulk
properties of the same nucleus from different studies to ensure that
they do refer to the same quantity. Trends of geometries with mass may
be equated nonetheless.

\begin{figure}
\scalebox{0.65}{\includegraphics*{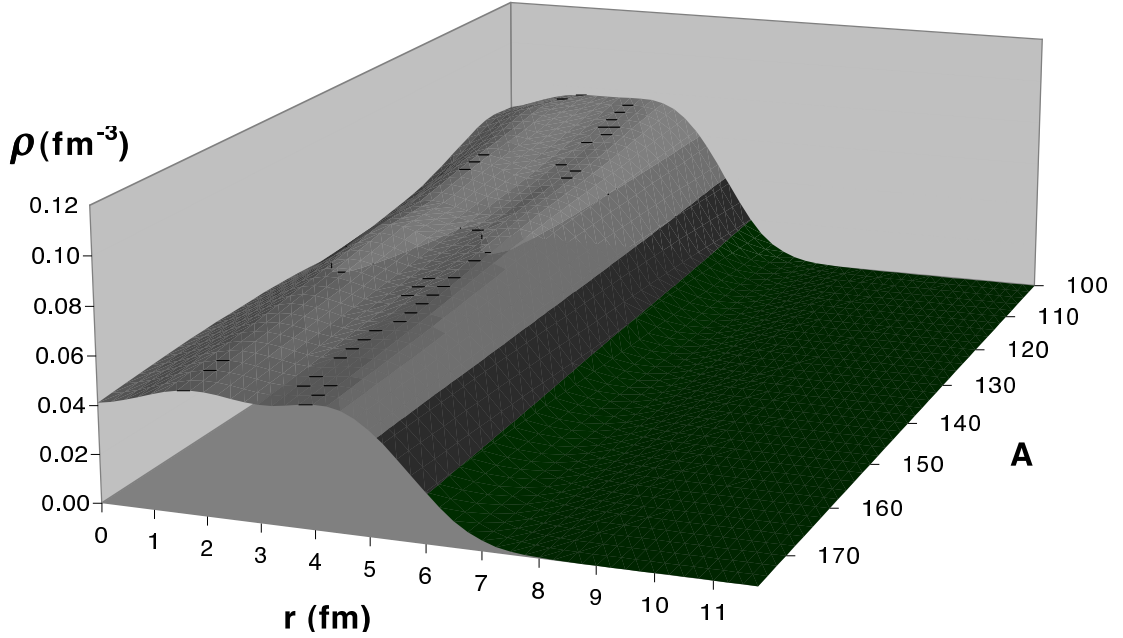}}
\caption{\label{Dens-prot-skl} The proton density variation 
with mass number from
the SLy4 model of structure for the Sn isotopes.}
\end{figure}

\begin{figure}
\scalebox{0.65}{\includegraphics*{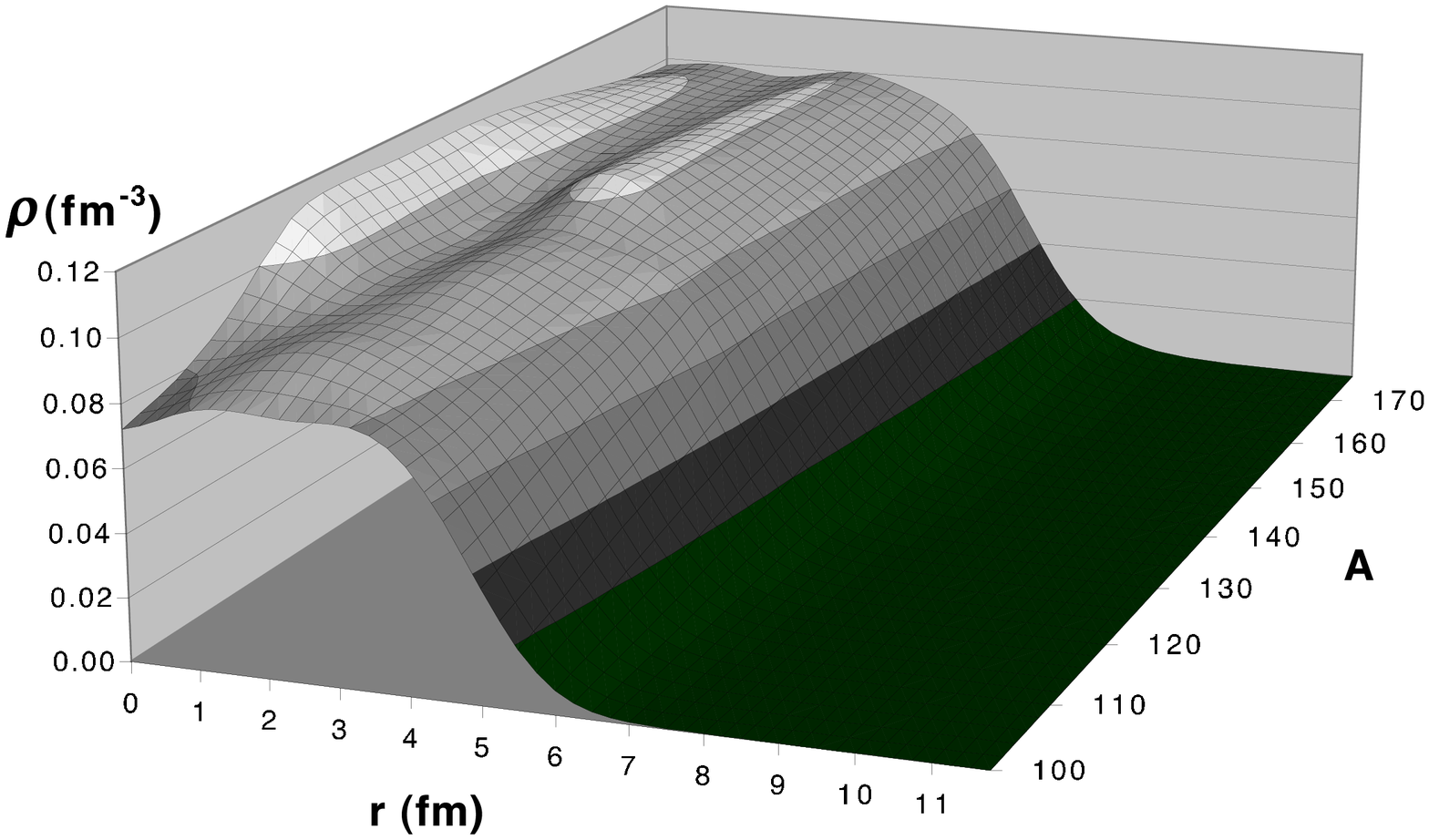}}
\caption{\label{Dens-neut-skl} The neutron density variation 
with mass number from
the SLy4 model of structure for the Sn isotopes.}
\end{figure}

\begin{figure}
\scalebox{0.65}{\includegraphics*{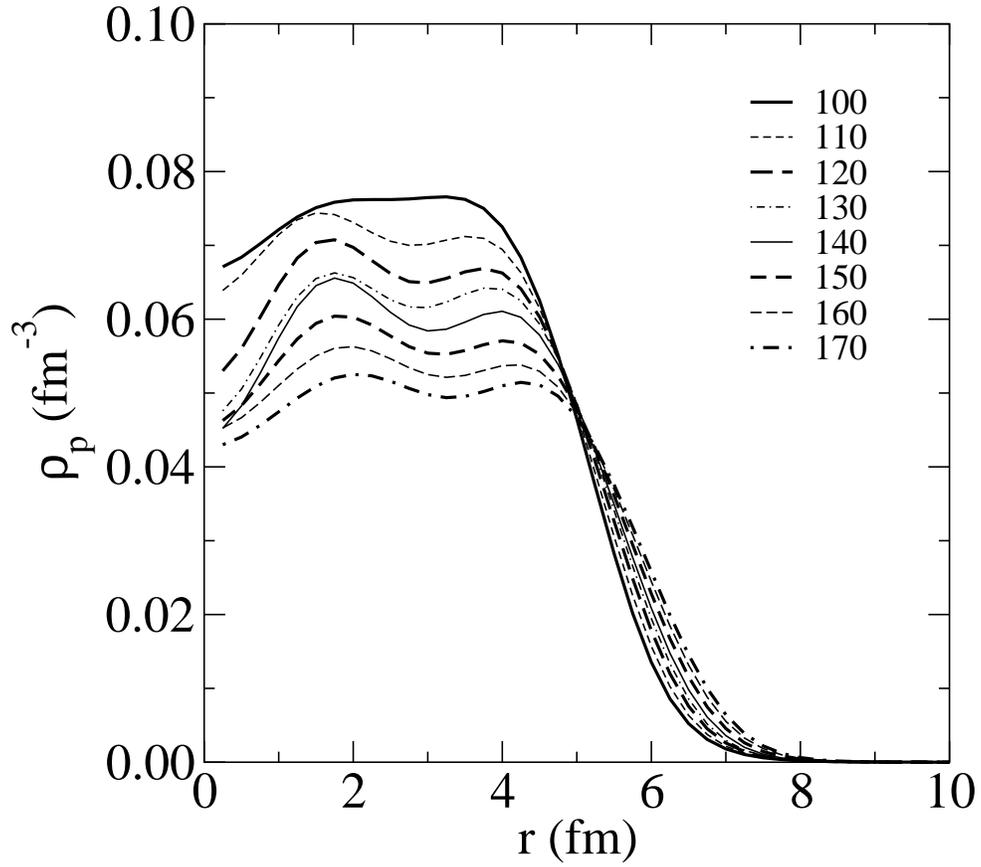}}
\caption{\label{flat-dens-prot-skl} The proton densities given by the
SkP model of structure for 8 isotopes of Sn.}
\end{figure}

\begin{figure}
\scalebox{0.65}{\includegraphics*{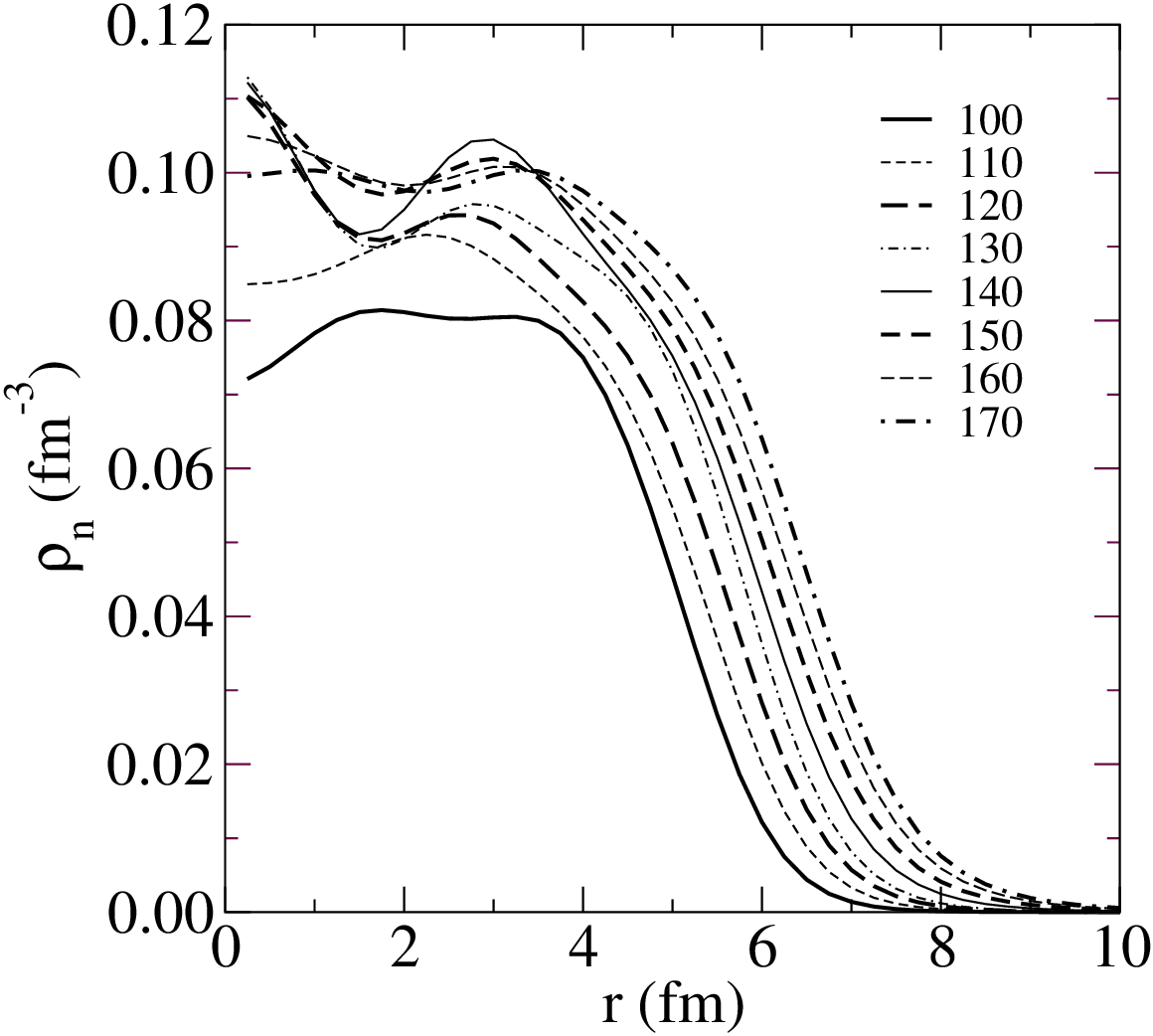}}
\caption{\label{flat-dens-neut-skl} The neutron densities given by the
SkP model of structure for 8 isotopes of Sn.}
\end{figure}

With increasing number of neutrons, from past~\cite{Mi00} as well as
these studies, the neutron and proton radii of the Sn isotopes
increase at different rates with the neutron radii increasing the
faster. As a result there is a gradual increase in the size of the
neutron skin; an increase that is almost linear with neutron number.
At the same time the neutron and proton bulk densities increase and
decrease, respectively. The balance between the bulk and surface
increase of the neutron distribution is governed by the volume and
surface attractions between neutrons and protons and hence is fixed by
the principal features of the volume and surface terms in nuclear
masses. That is evident from the actual density profiles deduced from
the shell occupancies and associated canonical wave functions of the
mean-field model results. Such complete density distributions for all
of the even mass Sn isotopes resulting from the SLy4 and SkP models of
their structure are so shown in Figs.~\ref{Dens-prot-skl} through
\ref{flat-dens-neut-skl}. The normalization of the neutron densities
we show is
\begin{equation}
4\pi \int_0^\infty \rho_n(r) \ r^2dr = N\ . 
\end{equation}
There is a similar form for the protons.

Using the SLy4 force, the proton and neutron matter distributions for
all even mass Sn isotopes resulting from the mean-field calculations
are displayed in Figs.~\ref{Dens-prot-skl} and \ref{Dens-neut-skl}
respectively. The proton (neutron) distributions are plotted with
nuclear mass decreasing (increasing) into the page. By that means the
variations to those densities are best portrayed in block form. In
these figures we display lines of equal radii and use different
shadings to designate regions over which the density changes by 0.02
fm$^{-3}$. The proton number of course is fixed at 50 and so as the
neutron number increases, and concomitantly the neutron volume
increases, those 50 protons extend over an increasing volume. As noted
above that is due to the strong attractive neutron-proton
interactions. In concert, the central charge density must, and does,
decrease. However, and as will be seen more clearly in
Fig.~\ref{flat-dens-prot-skl} below, the charge distributions do not
become significantly more diffuse. The prime effect is the $\sim 33\%$
increase in the charge volume.

The neutron densities structure variation with increasing mass is
quite different from that of the protons. Such is not unexpected as
the proton number is fixed at 50 with the neutron number increasing to
give the mass range. The general trend that the neutron rms radii
increase is evident as the half central density is reached at radii
ranging from $\sim 5$~fm in $^{100}$Sn to $\sim 6.5$~fm for
$^{170}$Sn. The increase of the neutron surface diffuseness is
observed readily in this figure by noting the changes along the lines
of fixed radius. We note also that a strong oscillation develops in
the central density, which on average also increases from $\sim
0.08$~neutrons/fm$^{3}$ in $^{100}$Sn to $\sim 0.1$~neutrons/fm$^{3}$
for $^{170}$Sn.

The mass variations of densities are evident also in
Figs.~\ref{flat-dens-prot-skl} and ~\ref{flat-dens-neut-skl} wherein
the proton and neutron densities respectively for the Sn isotopes
calculated using the SkP model are given for a select set of 8 nuclei
having masses spaced evenly between 100 and 170. With the SLy4 force
the mass trends are very similar with differences occurring in fine
details.

While a few of these densities have been depicted before~\cite{Mi00}
not only do we extend the set to a wider mass range and fill in
previous gaps but we highlight features that should most readily
relate to effects in predictions of scattering. In
Fig.~\ref{flat-dens-prot-skl} it is evident that the 50 protons are
rearranged to be more extensive as one increases mass. Note that the
half-density radius ranges from $\sim 5$~fm for $^{100}$Sn to $\sim
6$~fm in $^{170}$Sn. However the proton surface diffuseness remains
essentially unchanged. The distance over which the charge density
falls from 90\% to 10\% of its central value is $\sim 2$~fm in all
nuclei. That is also the case for the neutron distributions. There is
a gradual development of a neutron skin to the Sn isotopes, for while
with $^{100}$Sn the 50 protons and 50 neutrons have essentially the
same distribution (solid dark lines in the figures), the two density
profiles are somewhat disparate in $^{170}$Sn. Not only does the
neutron central density increase by $\sim 25\%$ from its value in
$^{100}$Sn while the proton central density value decreases by $\sim
40\%$, but the skin, in this case $R_{rms}\text{(neutron)} -
R_{rms}\text{(proton)}$, varies from 0 to $\sim 0.5$~fm as noted
previously from the definition in the Helm model
characterization~\cite{Mi00}.

It is of note also that the neutron density profiles have a mass
variation through the inner region ($\sim$ 4 fm) that is not as smooth
a progression as that in the outer radial region (beyond 4 fm). Those
changes in shape for masses in the vicinity of 150 we expect to
reflect as variations in cross section properties for momentum
transfer values $\sim$ 1 to 2 fm$^{-1}$ (at least for 200 MeV
scattering) as such did for analyses of data from 200 MeV proton
scattering from ${}^{208}$Pb~\cite{Ka02}.

\section{Mass variation of single particle level properties}

\subsection{The protons in the Sn isotopes}

\begin{figure}
\scalebox{0.65}{\includegraphics*{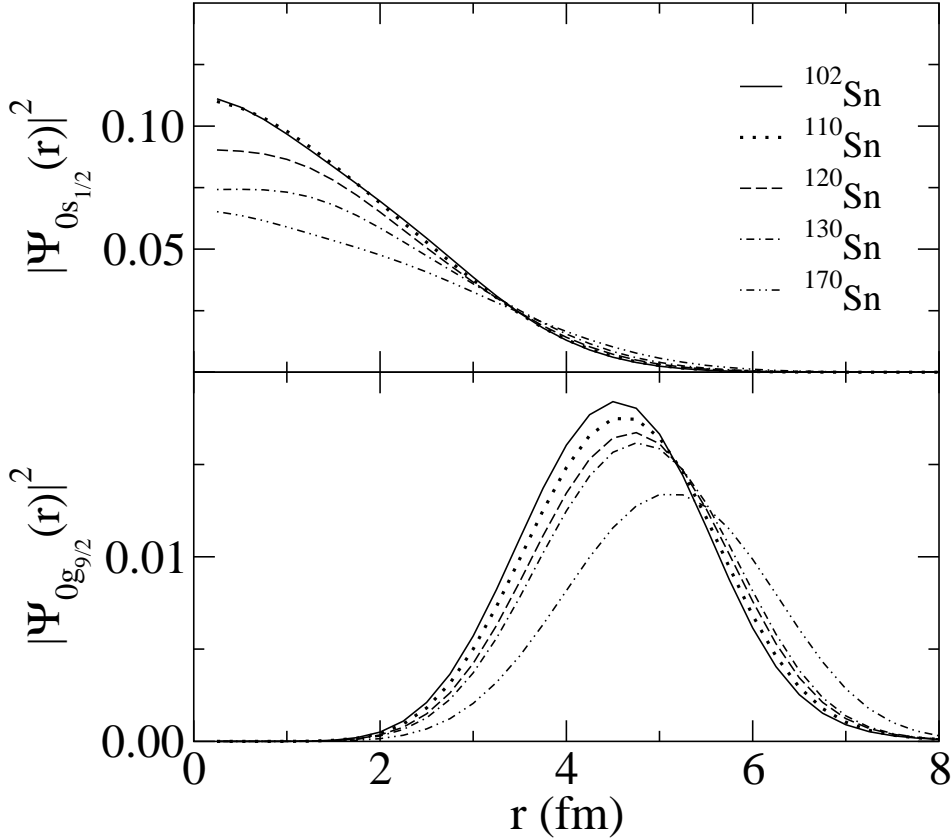}}
\caption{\label{Proton-fns}
The moduli square of the radial functions for the $0s_{\frac{1}{2}}$ (top) and 
$0g_{\frac{9}{2}}$ (bottom) proton orbits in ${}^{102,110,120,130,170}$Sn.}
\end{figure}

To a very high degree, the SLy4 and SkP models give the same closed orbit
proton configuration for the 50 protons in all of the Sn isotopes, i.e. 
\begin{equation}
(0s_{\frac{1}{2}})^2 (0p_{\frac{3}{2}})^4 (0p_{\frac{1}{2}})^2
(0d_{\frac{5}{2}})^6 (1s_{\frac{1}{2}})^2 (0d_{\frac{3}{2}})^4 
(0f_{\frac{7}{2}})^8 (1p_{\frac{3}{2}})^4 (0f_{\frac{5}{2}})^6 
(1p_{\frac{1}{2}})^2 (0g_{\frac{9}{2}})^{10}\ . 
\label{Coreconfig}
\end{equation}
However, the associated single proton wave functions vary with the mass 
of the isotope reflecting the expected expansion of the charge density
as the neutron numbers, and in concert the nuclear skin, increases. 
That variation is depicted in Fig.~\ref{Proton-fns}.  Therein are shown
the changes that occur over the mass range
in the modulus square of the  $0s_{\frac{1}{2}}$, the most bound of single
particle states, and in that of the $0g_{\frac{9}{2}}$ orbit, the latter being
a dominant element in defining the charge properties near and in the nuclear surface.
Clearly these wave functions extend as the neutron number increases, as do those of all 
of the other occupied orbitals.  But since the number of protons is fixed at 
50, the charge  density then gradually decreases in the nuclear center
while at long range that  becomes more diffuse.  This behavior, which is in stark
contrast to that of the neutron density, should be clearly evident if ever 
longitudinal form factors of these isotopes from electron scattering can be measured.

\subsection{The neutrons in the Sn isotopes}

Neutron orbit occupancies that result from the SLy4 
model calculations are presented in Tables~\ref{Sn102-130}, 
\ref{Sn134-158}, and \ref{Sn160-176}. They give the occupancies 
for orbits that are partially filled for each isotope of Sn. Orbits that
have occupancies for all the nuclei in each table that is less than 0.03 
are not listed.

Three of the isotopes have neutron orbits with complete occupancy.  They are 
${}^{100,132,176}$Sn of which only ${}^{176}$Sn is listed in the tables. The 
tables segment the mass range of the Sn isotopes with the neutron orbit 
occupancies for the set ${}^{102-130}$Sn given in Table~\ref{Sn102-130}. That 
ranges between the two isotopes having closed shell occupancies in the SLy4 
model. These lighter mass isotopes also have a neutron core basically 
the same as for the protons [Eq.~(\ref{Coreconfig})] though there is a small 
percentage breaking 
of full occupancy of the neutron $0g_{\frac{9}{2}}$ orbit as shown by the first
line in Table~\ref{Sn102-130}. 
\begin{table}
  \begin{ruledtabular}
\caption{\label{Sn102-130}
Valence orbit occupancies from the SLy4 model for the isotopes ${}^{102-130}$Sn.}
\begin{tabular}{cccccccccccccccc}
Orbit &\ \  102\ \  &\ \  104\ \  &\ \  106\ \  &\ \  108\ \  &\ \  110\ \  &\ \  112\ \  &
\ \  114\ \  &\ \  116\ \  &\ \  118\ \  &\ \  120\ \  &\ \  122\ \  &\ \  124\ \  &
\ \  126\ \  &\ \  128\ \  &\ \  130\ \  \\
\hline
$0g_{\frac{9}{2}}$  & 9.95 & 9.93 & 9.92 & 9.91 & 9.90 & 9.91 & 9.92 & 9.93 & 9.94 & 9.94 &
9.95 & 9.96 & 9.96 & 9.97 & 10.0 \\
$1d_{\frac{5}{2}}$  & 1.65 & 3.21 & 4.44 & 4.97 & 5.22 & 5.39 & 5.54 & 5.66 & 5.75 & 5.82 &
5.86 & 5.89 & 5.92 & 5.95 & 5.97 \\
$2s_{\frac{1}{2}}$  & 0.03 & 0.07 & 0.13 & 0.26 & 0.42 & 0.62 & 0.86 & 1.14 & 1.43 & 1.65 &
1.78 & 1.86 & 1.91 & 1.94 & 1.97 \\
$1d_{\frac{3}{2}}$  & 0.06 & 0.11 & 0.18 & 0.32 & 0.49 & 0.71 & 1.01 & 1.45 & 2.07 & 2.73 &
3.22 & 3.52 & 3.70 & 3.82 & 3.92 \\
$0g_{\frac{7}{2}}$  & 0.21 & 0.49 & 1.04 & 2.05 & 3.27 & 4.48 & 5.58 & 6.44 & 7.01 & 7.35 &
7.55 & 7.68 & 7.78 & 7.86 & 7.93 \\
$0h_{\frac{11}{2}}$ & 0.07 & 0.14 & 0.23 & 0.39 & 0.56 & 0.73 & 0.92 & 1.19 & 1.62 & 2.31 &
3.41 & 4.86 & 6.51 & 8.27 & 10.1\\
\hline
$0i_{\frac{13}{2}}$ &      & 0.02 & 0.02 & 0.03 & 0.04 & 0.04 & 0.05 & 0.05 & 0.05 & 0.05 &
0.05 & 0.05 & 0.05 & 0.04 & 0.03 \\
$1f_{\frac{7}{2}}$  &      & 0.01 & 0.02 & 0.02 & 0.03 & 0.03 & 0.04 & 0.04 & 0.04 & 0.04 &
0.05 & 0.05 & 0.05 & 0.04 & 0.03 \\ 
$0h_{\frac{9}{2}}$  &      & 0.01 & 0.02 & 0.02 & 0.03 & 0.04 & 0.04 & 0.04 & 0.04 & 0.04 &
0.05 & 0.05 & 0.05 & 0.04 & 0.03 \\
\end{tabular}
\end{ruledtabular}
\end{table}
As neutrons are added pairwise to ${}^{100}$Sn to reach mass 110, this model 
suggests that they primarily occupy the $1d_{\frac{5}{2}}$ orbit with some 
partial occupancy of both the $2s_{\frac{1}{2}}$ and  $0g_{\frac{7}{2}}$ orbits.
As mass then increases to reach ${}^{120}$Sn, the latter two orbits increase in
occupancy to 80-90\% while the $1d_{\frac{3}{2}}$ and $0h_{\frac{11}{2}}$ orbits
fill to  68\% and 20\% closure respectively and at approximately the same rate.
With the set of isotopes ${}^{122-132}$Sn, the major effect is in the filling of the
$0h_{\frac{11}{2}}$ orbit.  Higher shell states for all of these isotopes
account for less than a percent of the neutron numbers.  Of importance however,
is that, with the SLy4 model, filling of the four important valence orbits
is far from that one would guess with the simplest of shell schemes.

\begin{figure}
\scalebox{0.65}{\includegraphics*{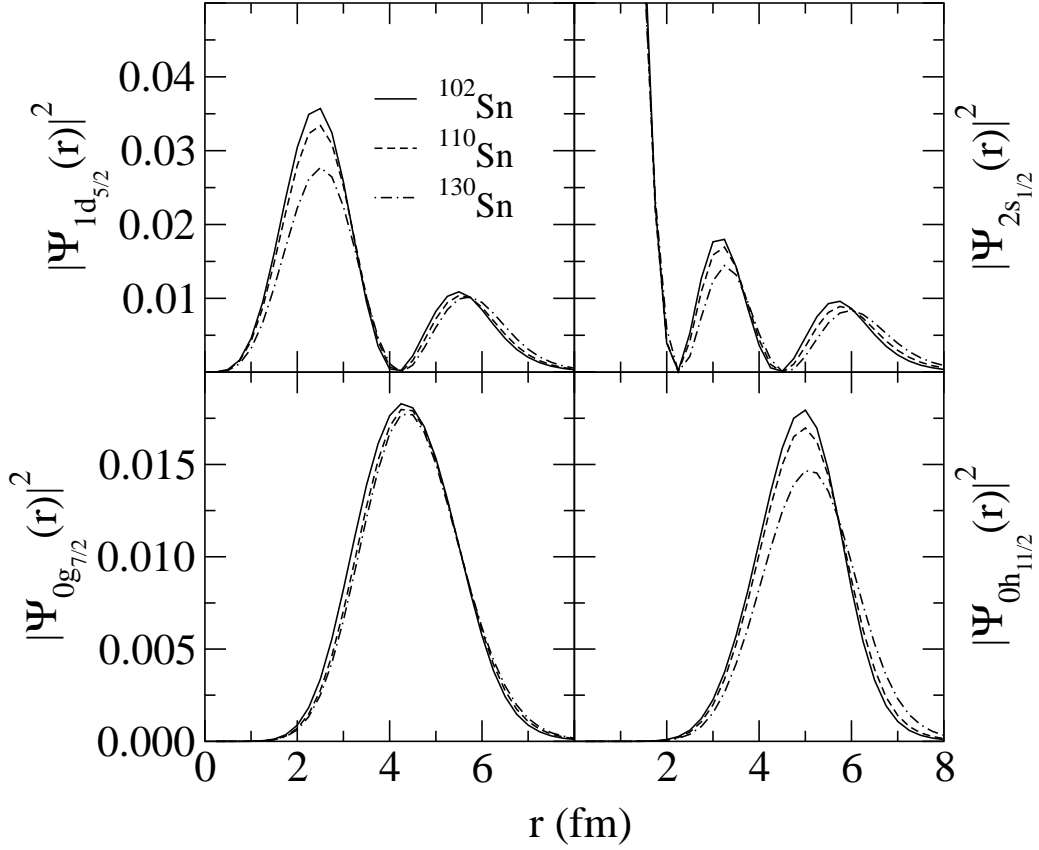}}
\caption{\label{Neutron1-fns}
The moduli square of the radial functions for four neutron orbits in 
${}^{102,110,130}$Sn.}
\end{figure}

But once again the valence orbitals vary as the 
mass increases.  That variation is illustrated in Fig.~\ref{Neutron1-fns} where
the four important valence orbital single particle densities are shown for three
isotopes, ${}^{102,110.130}$Sn. In this case we display just the changes wrought
between ${}^{102-130}$Sn since for this range of nuclei essentially those four
orbits are being filled with neutrons.  Across this range of isotopes each 
valence orbit gradually extends in space. The related neutron densities then 
increase in size as neutrons are added to form a neutron skin. But such a skin is not 
due simply to an increase in neutron occupancy of orbits with high angular momentum.
Each orbit's extension as the mass increases plays a role. 

In Tables~\ref{Sn134-158} and \ref{Sn160-176} respectively, 
the extra core neutron orbit occupancies for the isotopes ${}^{134-158}$Sn and 
${}^{160-176}$Sn are given.  For all of those isotopes the neutron core essentially
has 62 neutrons.  The 12 neutrons extra to the 50 in the configuration of 
Eq.~(\ref{Coreconfig})
closes the $2s$-$1d$ shell.
\begin{table}
  \begin{ruledtabular}
\caption{\label{Sn134-158}
Valence orbit occupancies from the SLy4 model for the isotopes ${}^{134-158}$Sn.}
\begin{tabular}{cccccccccccccc}
Orbit &\ \  134\ \  &\ \  136\ \  &\ \  138\ \  &\ \  140\ \  &\ \  142\ \  &
\ \  144\ \  &\ \  146\ \  &\ \  148\ \  &\ \  150\ \  &\ \  152\ \  &\ \  154\ \  &
\ \  156\ \  &\ \  158\ \  \\
\hline
$0g_{\frac{7}{2}}$  & 8.00 & 8.00 & 8.00 & 8.00 & 8.00 & 7.98 & 7.97 &
 7.97 & 7.97 & 7.97 & 7.97 & 7.97 & 7.97 \\
$0h_{\frac{11}{2}}$ & 11.96 & 11.95 & 11.96 & 11.98 & 11.93 & 11.90 & 
 11.89 & 11.88 & 11.87 & 11.87 & 11.88 & 11.88 & 11.89 \\
$1f_{\frac{7}{2}}$  & 1.82 & 3.67 & 5.57 & 7.62 & 7.98 & 7.40 & 7.41 & 
 7.44 & 7.47 & 7.52 & 7.57 & 7.62 & 7.67 \\
$0h_{\frac{9}{2}}$  & 0.07 & 0.14 & 0.17 & 0.15 & 0.97 & 1.78 & 2.63 & 
 3.51 & 4.40 & 5.28 & 6.12 & 6.88 & 7.56 \\
$0i_{\frac{13}{2}}$ & 0.05 & 0.09 & 0.11 & 0.07 & 0.37 & 0.63 & 0.88 & 
 1.13 & 1.39 & 1.66 & 1.96 & 2.31 & 2.71 \\
$1f_{\frac{5}{2}}$  & 0.04 & 0.07 & 0.08 & 0.06 & 0.35 & 0.63 & 0.92 & 
 1.23 & 1.56 & 1.94 & 2.35 & 2.81 & 3.31 \\
$2p_{\frac{3}{2}}$  & 0.03 & 0.05 & 0.07 & 0.09 & 0.73 & 1.31 & 1.80 & 
 2.19 & 2.51 & 2.77 & 3.00 & 3.19 & 3.36 \\
$2p_{\frac{1}{2}}$  &      & 0.01 & 0.02 & 0.02 & 0.10 & 0.19 & 0.28 & 
 0.38 & 0.50 & 0.63 & 0.77 & 0.93 & 1.11 \\
\hline
$1g_{\frac{9}{2}}$  &      & 0.01 & 0.01 & 0.03 & 0.03 & 0.04 & 0.05 & 
 0.06 & 0.07 & 0.08 & 0.09 & 0.09 & 0.10 \\
$0j_{\frac{15}{2}}$ &      & 0.01 & 0.01 & 0.03 & 0.03 & 0.04 & 0.05 & 
 0.06 & 0.07 & 0.08 & 0.08 & 0.08 & 0.08 \\
$0i_{\frac{11}{2}}$ &      &      &      &      & 0.02 & 0.03 & 0.04 & 
 0.05 & 0.06 & 0.06 & 0.07 & 0.07 & 0.07 \\
$2d_{\frac{5}{2}}$  &      &      &      &      &      & 0.01 & 0.02 & 
 0.02 & 0.02 & 0.02 & 0.02 & 0.03 & 0.03 \\
$1g_{\frac{7}{2}}$ &       &      &      &      &      & 0.01 & 0.02 & 
 0.02 & 0.03 & 0.03 & 0.03 & 0.03 & 0.03 \\
$0k_{\frac{17}{2}}$ &      &      &      &      &      & 0.01 & 0.02 & 
 0.02 & 0.02 & 0.02 & 0.03 & 0.03 & 0.03 \\
\end{tabular}
\end{ruledtabular}
\end{table}

The Sly4 model neutron orbit occupancies for the isotopes ${}^{134- 158}$Sn
given in Table~\ref{Sn134-158} reveal  that the $0g_{\frac{7}{2}}$ and 
$0h_{\frac{11}{2}}$ orbits are almost fully occupied. Thus in adding neutrons to
reach ${}^{142}$Sn, those neutrons almost all go into the $1f_{\frac{7}{2}}$ orbit, 
almost filling it completely. As neutrons are added
thereafter, that $1f_{\frac{7}{2}}$ orbit occupancy decreases slightly before slowly
regaining to almost full occupancy in ${}^{160}$Sn. Further, in changing from
mass 144 to mass 162, the other orbits increase in occupancy, more or less at the same
rate to result in 90\% full for the $0h_{\frac{9}{2}}$ and $2p_{\frac{3}{2}}$ orbits,
70\% full for the $1f_{\frac{5}{2}}$ and $2p_{\frac{1}{2}}$ orbits, and 30\%
occupation for the $0i_{\frac{13}{2}}$ orbit.  Higher shells to these again account
for less than 1\% of the neutron numbers.

\begin{figure}
\scalebox{0.65}{\includegraphics*{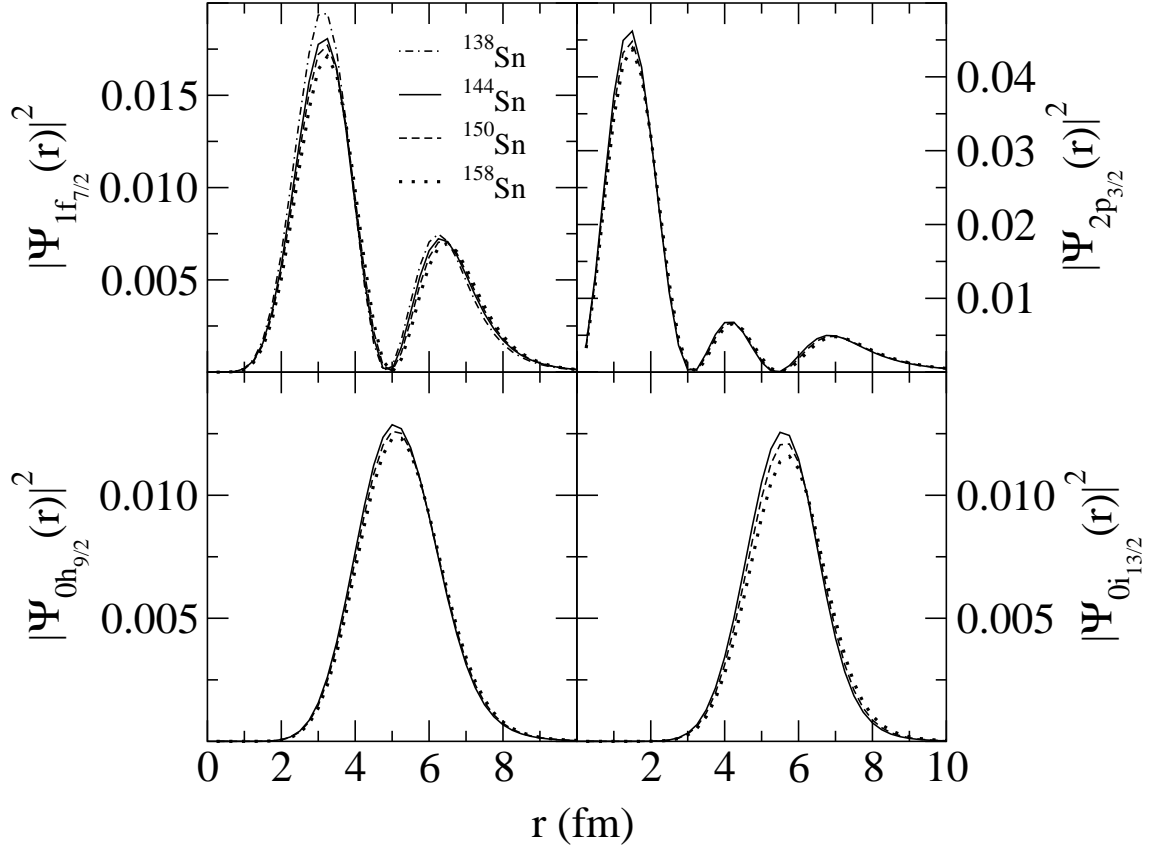}}
\caption{\label{Neutron2-fns}
Radial wave functions for four neutron orbits
in ${}^{134,144,150,158}$Sn.}
\end{figure}

As before, as one moves across this range of isotopes, the individual orbit wave
functions for the neutrons become more extended.  That is shown in Fig.~\ref{Neutron2-fns}
wherein four (valence) orbit functions for ${}^{134,144,150,158}$Sn are displayed.
For ${}^{138}$Sn, only the $1f_{\frac{7}{2}}$ wave function is shown (by the 
dot-dashed curve) since that is the only valence orbit of the four in that nucleus with
a substantial neutron occupancy. However, for all four orbitals, the wave functions
determined from the SLy4 calculations are portrayed by the solid, dashed, and
dotted curves for $^{144}$Sn, $^{150}$Sn, and $^{158}$Sn respectively.
Again these valence orbit wave functions extend to larger radii as the isotope mass
increases, but now not as markedly as for the protons or neutron orbits for the 
lighter mass isotopes.  It follows then that the build up of neutron density at the 
nuclear surface is more influenced now by the increasing occupancies of these orbits.

\begin{table}
  \begin{ruledtabular}
\caption{\label{Sn160-176}
Valence orbit occupancies from the SLy4 model for the isotopes $^{160-176}$Sn.}
\begin{tabular}{cccccccccc}
Orbit &\ \ 160 \ \ &\ \  162\ \ &\ \  164\ \  &\ \  166\ \  &\ \  168\ \  &\ \  170\ \  &\ \  172\ \  &
\ \  174\ \  &\ \  176\ \ \\
\hline
$0g_{\frac{7}{2}}$  & 7.97 & 7.98 & 7.98 & 7.98 & 7.98 & 8.00 & 8.00 & 8.00 & 8.00 \\
$0h_{\frac{11}{2}}$ & 11.90 & 11.91 & 11.92 & 11.93 & 11.94 & 11.96 & 11.97 & 11.98 & 12.00 \\ 
$1f_{\frac{7}{2}}$  & 7.72 & 7.77 & 7.81 & 7.84 & 5.36 & 7.91 & 7.94 & 7.97 & 8.00 \\ 
$0h_{\frac{9}{2}}$  & 8.13 & 8.59 & 8.96 & 9.24 & 9.46 & 9.63 & 9.77 & 9.89 & 10.00 \\ 
$0i_{\frac{13}{2}}$ & 3.23 & 3.92 & 4.82 & 5.98 & 7.35 & 8.87 & 10.50 & 12.21 & 14.00 \\ 
$1f_{\frac{5}{2}}$  & 3.82 & 4.32 & 4.76 & 5.10 & 5.36 & 5.56 & 5.72 & 5.87 & 6.00 \\ 
$2p_{\frac{3}{2}}$  & 3.50 & 3.62 & 3.71 & 3.78 & 3.84 & 3.88 & 3.92 & 3.96 & 4.00 \\ 
$2p_{\frac{1}{2}}$  & 1.30 & 1.48 & 1.61 & 1.73 & 1.81 & 1.87 & 1.92 & 1.96 & 2.00 \\
\hline
$1g_{\frac{9}{2}}$  & 0.10 & 0.10 & 0.10 & 0.10 & 0.10 & 0.09 & 0.07 & 0.04 & --- \\
$0j_{\frac{15}{2}}$ & 0.08 & 0.08 & 0.08 & 0.08 & 0.08 & 0.07 & 0.05 & 0.03 & --- \\ 
$0i_{\frac{11}{2}}$ & 0.07 & 0.07 & 0.07 & 0.07 & 0.07 & 0.06 & 0.05 & 0.03 & --- \\ 
$2d_{\frac{5}{2}}$  & 0.03 & 0.03 & 0.02 & 0.02 & 0.02 & 0.02 & 0.02 & --- & --- \\ 
$1g_{\frac{7}{2}}$  & 0.03 & 0.03 & 0.03 & 0.03 & 0.03 & 0.02 & 0.02 & 0.01 & --- \\ 
$0k_{\frac{17}{2}}$ & 0.03 & 0.03 & 0.02 & 0.02 & 0.02 & 0.02 & 0.01 & --- & --- \\
\end{tabular}
\end{ruledtabular}
\end{table}

The neutron orbit occupancies given by the SLy4 model for the heaviest Sn isotopes,
masses 160 to 176, are given in Table~\ref{Sn160-176}. Those trend to a complete shell 
occupancy for ${}^{176}$Sn.  The most obvious change in occupancy over this set of nuclei
is that of the $0i_{\frac{13}{2}}$ orbit (from 3.23 to 14 neutrons). Again the higher 
orbits are sparsely occupied, and in fact any such occupancy decreases until 
${}^{176}$Sn adopts the closed shell character. 

\begin{figure}
\scalebox{0.65}{\includegraphics*{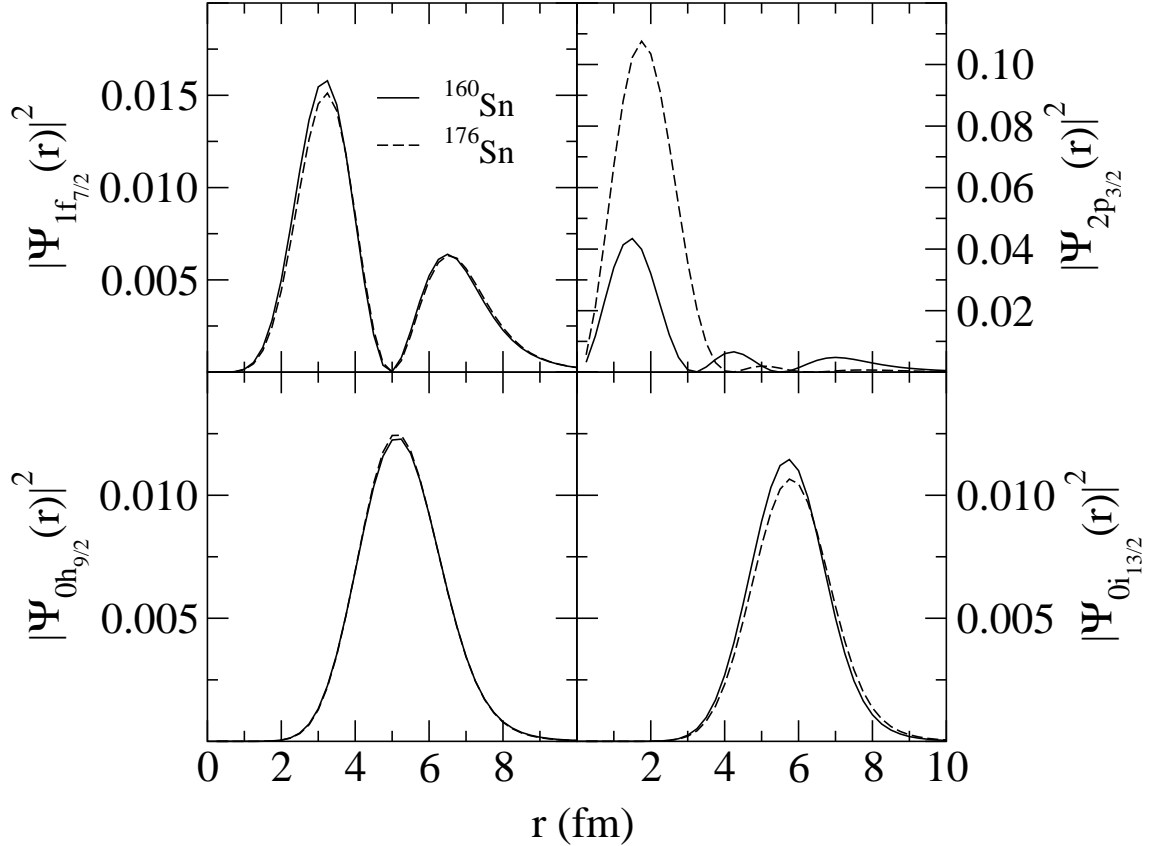}}
\caption{\label{Neutron3-fns}
Radial wave functions for four neutron orbits
in ${}^{160}$Sn and ${}^{176}$Sn.}
\end{figure}
The wave functions for the valence orbitals of importance for the heaviest Sn isotopes
are displayed in Fig.~\ref{Neutron3-fns} where the solid curve are the functions
specified for ${}^{160}$Sn and the dashed curves show those for ${}^{176}$Sn.


\section{Scattering from Hydrogen -- gedanken results}
\label{res-scat}

We have used the canonical SP wave functions for all of
the Sn isotopes to fold an effective $NN$ interaction and thus
generate optical potentials for use in modified versions of the
scattering program DWBA98~\cite{Ra98}. Those optical potentials are
complex and non-local since they can be written in coordinate space
with $\bm{r}_{12} = \bm{r}_1 - \bm{r}_2$, as
\begin{equation}
U\left( \bm{r}_1, \bm{r}_2; E \right) = U_D \left( \bm{r}_1; E
\right) \delta\left( \bm{r}_{12} \right) + U_E\left( \bm{r}_1,
\bm{r}_2; E \right) \, ,
\label{NonSE}
\end{equation}
where the direct $U_D$ and non-local exchange $U_E$ terms are
\begin{eqnarray}
U_D & = & \sum_n \zeta_n \left\{ \int \varphi^{\ast}_n( \bm{s} )
v_D(\bm{r}_{1s}) \varphi_n( \bm{s} ) \, d\bm{s} \right\} \, \nonumber \\
U_E & = & \sum_n \zeta_n \left\{ \varphi^{\ast}_n( \bm{r}_1 ) v_{Ex}(
\bm{r}_{12} ) \varphi_n( \bm{r}_2 ) \right\} \, .
\end{eqnarray}
Here $v_D$ and $v_{Ex}$ are combinations of the components of the effective 
$NN$ interactions, $\zeta_n$ are ground state OBDME, and $\varphi_n(\bm{s})$ 
are nucleon SP states.  The OBDME are represented as bound state shell 
occupancies.  All details and the prescription of solution of the associated 
nonlocal Schr\"odinger equations are given in the recent review~\cite{Am00}.  
The results to be discussed herein have been obtained by solving the actual 
nonlocal Schr\"odinger equations defined with potentials as given by
Eq.~(\ref{NonSE}). For the present calculations, the effective $NN$ 
interactions have been defined by their mapping to the BBG $g$ matrices of 
the BonnB $NN$ interaction~\cite{Ma87}.
  
We consider the elastic scattering of 200~MeV protons from each of the
even mass Sn isotopes ($A = 100$ to 176).  By inverse kinematics the
cross sections we determine are also those for the scattering of
$200A$~MeV Sn ions from hydrogen; some of which can be, or may soon
be, obtained in sufficient numbers to form a radioactive ion beam for
experiment.  The choice of 200~MeV was made, not only because the
effective force at this energy has been used in many successful
predictions of actual scattering data from stable nuclei~\cite{Am00}
but also  data at that energy from $^{208}$Pb gave a
clear signature of its neutron density profile~\cite{Ka02}.

\begin{figure}
\scalebox{0.65}{\includegraphics*{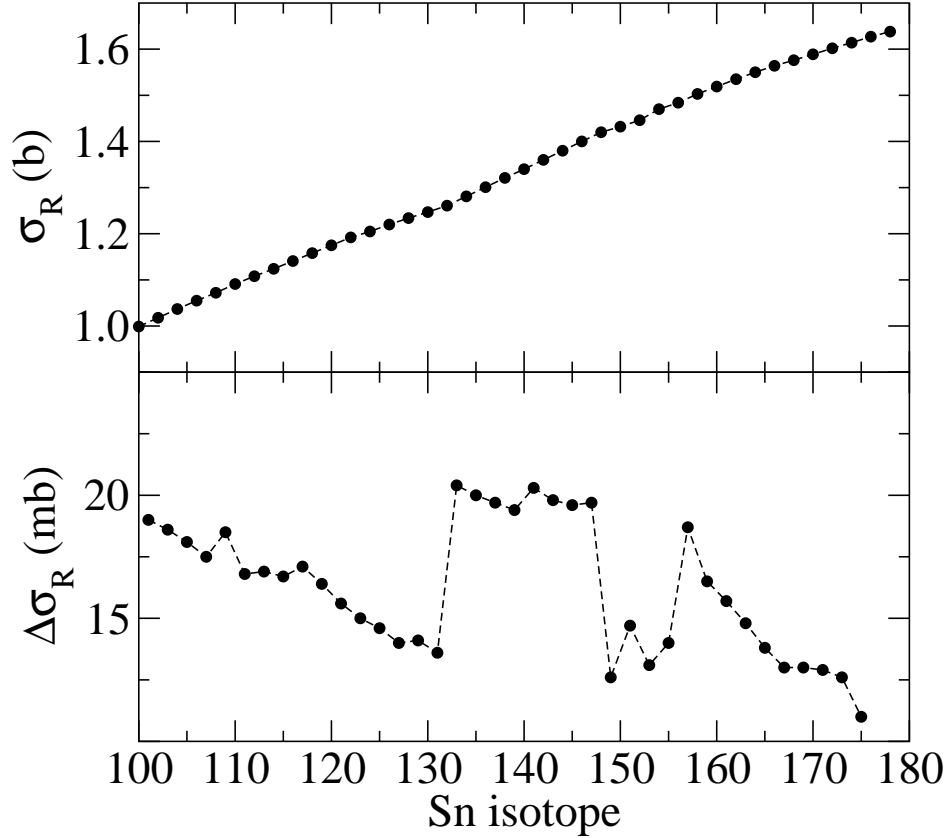}}
\caption{\label{Sn-sigr-diffs} 
Differences of the total reaction cross sections for 200 MeV proton scattering
from adjacent even mass Sn isotopes.}
\end{figure}

Possibly first measurements with such exotic nuclear projectiles  may be of 
the total reaction cross sections and the SLy4 model predictions for those 
are shown  in the top panel of Fig~\ref{Sn-sigr-diffs}.  There is a distinct a mass 
trend in those results so that total reaction cross section  gradually 
increases with isotope mass from $\sim$ 1 b to 1.63 b 
for ${}^{170}$Sn.  The sums that define the reaction cross sections are 
dominated by the large partial wave scattering amplitudes~\cite{De02}.  Also, 
at much higher energies, those sums are known to equate to the geometric cross 
section for each nucleus.  So reaction cross section values reflect a bulk 
character of the structure of each nucleus and  values of the total reaction cross 
sections may then be the first obvious evidence of the propriety of any chosen 
model for the structure of the isotopes.
If any such can be measured accurately the differences also may then be a first 
test of the structure models since, as portrayed in the bottom segment of 
Fig.~\ref{Sn-sigr-diffs}, there
is a noticeable variation of many mb predicted when the SLy4 model 
structures are used in calculations of the scattering of 200 MeV protons.
Notably that the differences show a trend to smaller values with increasing
mass save for a ``quantum" jump when the neutron number is 82 and for some uniqueness
with masses 150 to 156. 
But any such clear variation will require an accuracy of measurement 
of a percent or two.
Nonetheless the actual value first, and the trend of differences second,
may signify propriety of the structures chosen to describe the isotopes.

\begin{figure}
\scalebox{0.65}{\includegraphics*{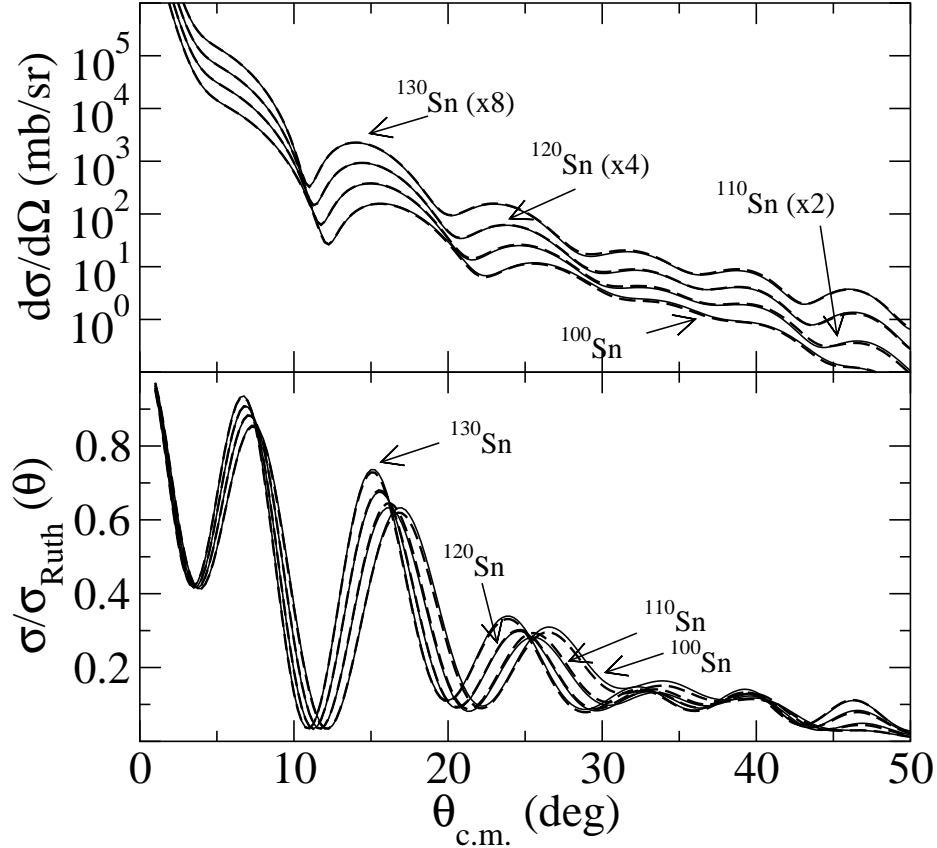}}
\caption{\label{Sn-200-cs+RR} The cross sections  from 200 MeV proton
scattering with select Sn isotopes. The differential cross sections
are shown in the to panel while the ratios to Rutherford results
are displayed in the bottom panel.}  
\end{figure}

\begin{figure}
\scalebox{0.65}{\includegraphics*{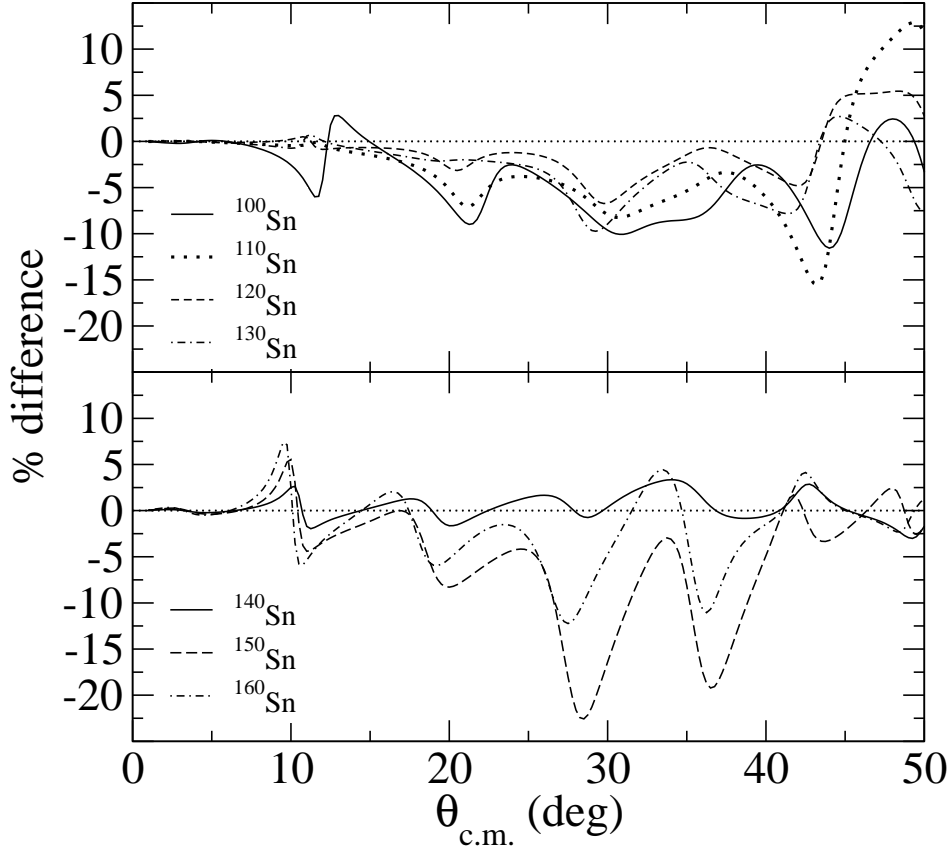}}
\caption{\label{SKL-SKP-diff} 
Percentage differences between differential cross sections for 200 MeV proton 
scattering and calculated using the two structure model densities.}
\end{figure}

\begin{figure}
\scalebox{0.65}{\includegraphics*{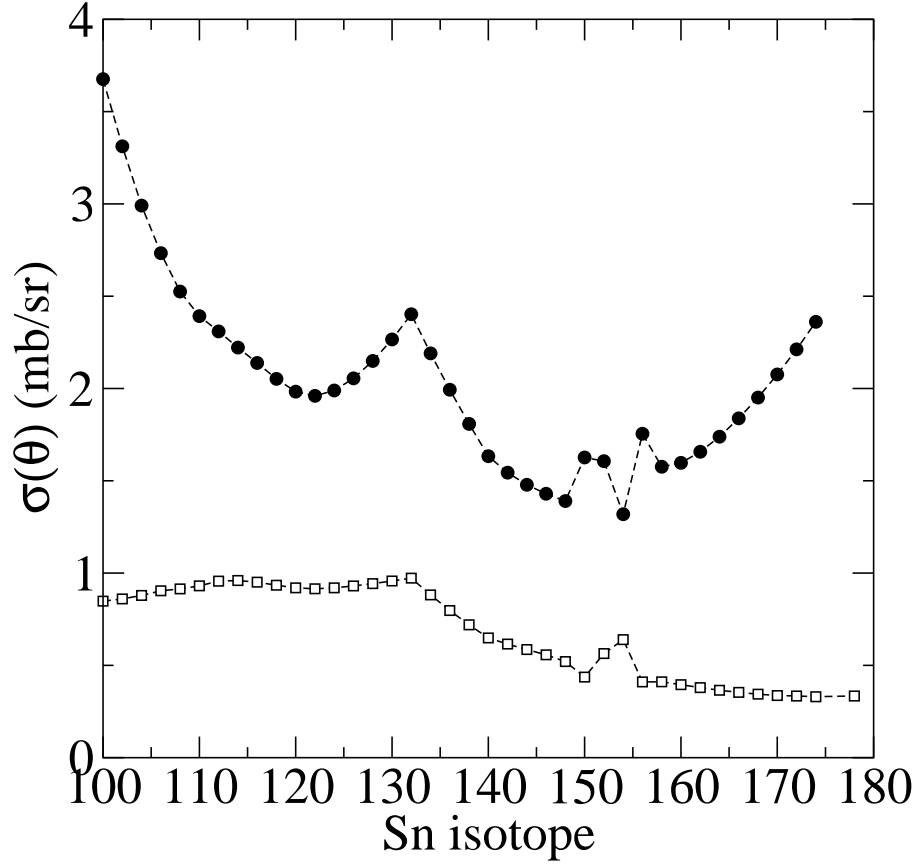}}
\caption{\label{Sn-30-40-set} The differential cross section values at
$30^\circ$ (circles) and $40^\circ$ (squares) scattering for all even
mass Sn isotopes as evaluated using the SLy4 model of their
structure.}
\end{figure}

There appears to be little difference  between the cross sections found 
using the SkP and the SLy4 forces to give the ground state structures.  This
is shown for a select set of Sn isotopes in Fig.~\ref{Sn-200-cs+RR}.
The differential cross sections for 200 MeV proton scattering from 
${}^{100,110,120,130}$Sn are shown in the top panel with each scaled by 
factors of 1, 2, 4, and 8 to distinguish them from one another more clearly.
The ratio to Rutherford scattering for these same nuclei and at 200 MeV are
displayed in the bottom panel of this figure.  The solid (dashed)
curves were obtained by using the structure found with the SkP (SLy4)
force. On this scale there is little to distinguish between the 
alternate model results.  But the two structure models do predict cross 
sections that differ in detail and such is shown in Fig.~\ref{SKL-SKP-diff} 
for the set of isotopes ${}^{100,110,120,130,140,150,160}$Sn. Therein 
the percentage difference,
\begin{equation}
\%\ {\rm difference}\ = 
\left[ \frac{d\sigma_{SkP}}{d\Omega}- \frac{d\sigma_{SLy4}}{d\Omega}\right]/
{\left|\frac{d\sigma_{SkP}}{d\Omega}\right|}\times 100
\label{percdiff}
\end{equation}
is plotted for each nucleus as identified.
Differences can be large (20\%) as with ${}^{150}$Sn coinciding with cross 
section magnitudes of a few mb/sr  at scattering angles $\sim 30 - 40^\circ$. 
That coincides with momentum transfer values of $\sim 1$ to 2 fm$^{-1}$ and 
are values which might be distinguished by measurement.  In the main however, 
the model differences characteristically are 5 to 10 \%, so that at scattering 
angles of 20$^\circ$ where predicted cross sections (for 200 MeV) are about 20
mb/sr, the difference between the model predictions would be about a mb/sr. 
For larger angle results, where the cross section magnitude is of the order of 
a mb/sr the differences, $\sim 0.05 - 0.1$ mb/sr may not exceed experimental 
uncertainties.  Nonetheless there is a progression in these cross section 
differences found using the sets of matter densities given by the SkP and 
SLy4 structure models as one increases the isotope mass and when compared 
with data, such might suggest  preference for one model of structure over the 
other; presupposing of course that one is indeed a better description of 
reality. To illustrate this further, in Fig.~\ref{Sn-30-40-set} we show the 
cross section values at 30$^\circ$  and 40$^\circ$ scattering that result from 
using the SLy4 structure model to form
$g$-folding optical potentials. The momentum transfers for these scattering
angles are $\sim 1.5$ and $\sim 2$ fm$^{-1}$ respectively. At these momentum
transfer values, clearly the variation of the structure details with mass have 
noticeable effect. There is some evidence for a more packed neutron distribution
in the structures at neutron number 82 and some evidence of surface orbit
variations with neutron numbers 100, 102, and 104. The latter, however, may simply
be an idiosyncrasy of the model calculations.
Thus, and as found in a study of scattering from ${}^{208}$Pb~\cite{Ka02},
the neutron distribution through the nuclear surface in particular, reflects
strongly in the calculated values for elastic scattering cross sections
for momentum transfers 1 to 3 fm$^{-1}$ with cross section values $\sim 1$ 
mb/sr.

\begin{figure}
\scalebox{0.7}{\includegraphics*{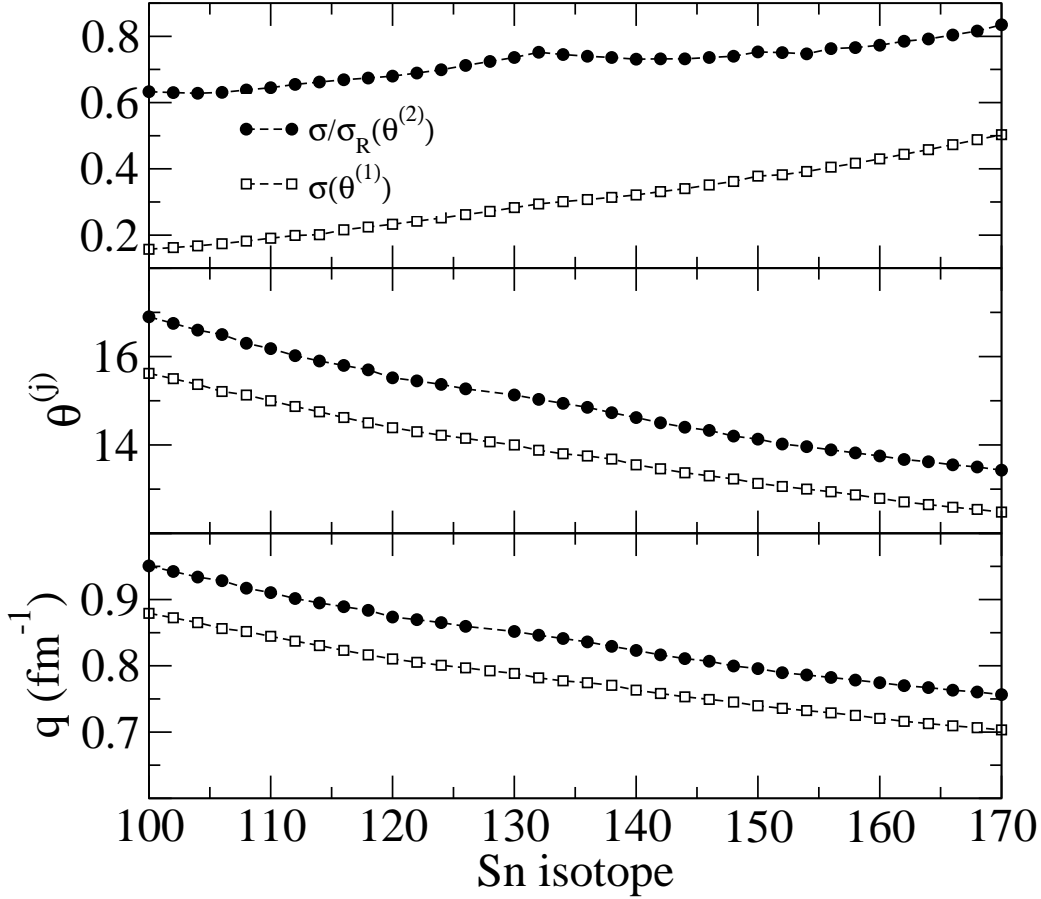}}
\caption{\label{Peak-xsec} 
Some low momentum transfer characteristics of 200 MeV proton scattering 
from the 110-170 even mass Sn isotopes. The results for $j=1$ (see
text) are denoted by the circles while those for $j=2$ are denoted by the
squares.}
\end{figure}

Other properties of the cross sections reflect possible measurable trends.
In Fig.~\ref{Peak-xsec} diverse variations with mass in the entire set of 
isotopes are shown. Such are quantities from scattering having analogies with 
the bulk nuclear properties determined previously from the mean-field matter 
densities~\cite{Mi00}.
In the top panel the variations with mass of magnitudes of select (low 
momentum) peaks in the calculated cross sections are shown. The first
obvious peak in the differential cross section is displayed by the filled 
circles. That changes markedly as one progresses from proton to neutron drip 
lines, from the value of $\sim 150$~mb/sr for ${}^{100}$Sn to a value of 
$\sim 500$ mb/sr for ${}^{170}$Sn as shown and on to $\sim 540$ mb/sr for 
${}^{176}$Sn. As with the total reaction cross section results this variation 
would be a measurable effect in future scattering experiments, and one that 
may be an easier characteristic to define than the higher momentum properties
we described earlier. The $y$-axis in this case is to be read as b/sr.
In the top panel we also display the values of the ratio to Rutherford cross 
sections found at the scattering angles corresponding to the second peak in 
those ratios.  They are portrayed by the opaque squares with the y-axis now in 
dimensionless units.  Again there is a smooth measurable effect with mass, both
in the magnitude of that second ratio to Rutherford peak as well as with the 
angles at which that peak occurs. We have designated these trends as low 
momentum properties as specified by the prescriptions. The particular 
values involved are shown in the middle and bottom panels of this figure.  In 
the middle panel the center of mass scattering angles at which the differential 
cross sections have their first definite peak are shown by the filled circles 
while those at which the ratio to Rutherford cross sections have their second 
peak are displayed once again by the opaque squares.  With increasing mass, 
and for all the isotopes, the first prominent 
peak of the differential cross sections move inward to smaller values of 
momentum transfer. Over the entire range, the momentum transfer value at which 
the first peak occurs changes from 0.88 fm$^{-1}$ in ${}^{100}$Sn to 0.7 
fm$^{-1}$ in ${}^{170}$Sn.

\section{Comparison with data}
\label{res-data}

There have been many experimental studies of proton scattering from 
Sn isotopes.  Our literature search on those gave the listing as shown in 
Table~\ref{Sntable}.  The proton energy was restricted to the range 16 to 200
MeV; a range for which we felt most comfortable about using the $g$-folding
method of defining optical potentials, and for which range we are confident about
the effective $NN$ interactions we have defined~\cite{Am00}.
\begin{table}
\begin{ruledtabular}
\caption{\label{Sntable}
Selected list of references to data and analyses of data
from proton elastic scattering from Sn isotopes.}
\begin{tabular}{ccl}
  Energy & Isotopes & References\\
  \hline
16.0 & 116, 120, 124 & \cite{Bo71}, \cite{Ma71}, \cite{Ab87} \\ 
20.4 & 116, 118, 120, 122, 124 & \cite{Wa89} \\
24.5 & 116, 118, 120, 124 & \cite{Ta81}, \cite{Pi85}, \cite{Wa89} \\
30.3 & 112, 114, 116, 118, 120, 122, 124 &
\cite{Cr64}, \cite{Ge70}, \cite{Hn71}, \cite{Ha71}, \cite{Ma75} \\
39.6 & 116, 118, 120, 122, 124 & \cite{Fr67}, \cite{Bo68}, \cite{Ha75} \\ 
49.35 & 112, 114, 118, 120, 124 & \cite{Ma71} \\
61.4 - 65 & 118, 120 & \cite{Fu69}, \cite{Ha80}, \cite{Ma88} \\
100.0 - 104.0 & 120 & \cite{Ha75}, \cite{Kw78}, \cite{Sc82}, \cite{Ka84} \\
133.8 & 116 & \cite{We87} \\
156.0 - 160.0 & 116, 118, 120 & \cite{Co74}, \cite{Sa02} \\
200 & 120 & \cite{Ka01} 
\end{tabular}
\end{ruledtabular}
\end{table}

From that list we have chosen for analysis, some of the data taken at 39.8, 
49.35, 65.0, and 200.0 MeV. A complete analysis of all of the available data
will be made and reported in some fashion subsequently. In all of the figures
to be shown, unless otherwise stated, the calculated results were obtained
from optical potentials formed by folding a Melbourne effective $NN$ 
interaction~\cite{Am00} for the relevant incident energy, with the OBDME and 
SP wave functions determined by the SLy4 model of structure for each
and every isotope of Sn. The DWBA98 program~\cite{Ra98} was used to perform
the folding and to find solutions of the relevant Schr\'odinger 
equations.

\begin{figure}
\scalebox{0.65}{\includegraphics*{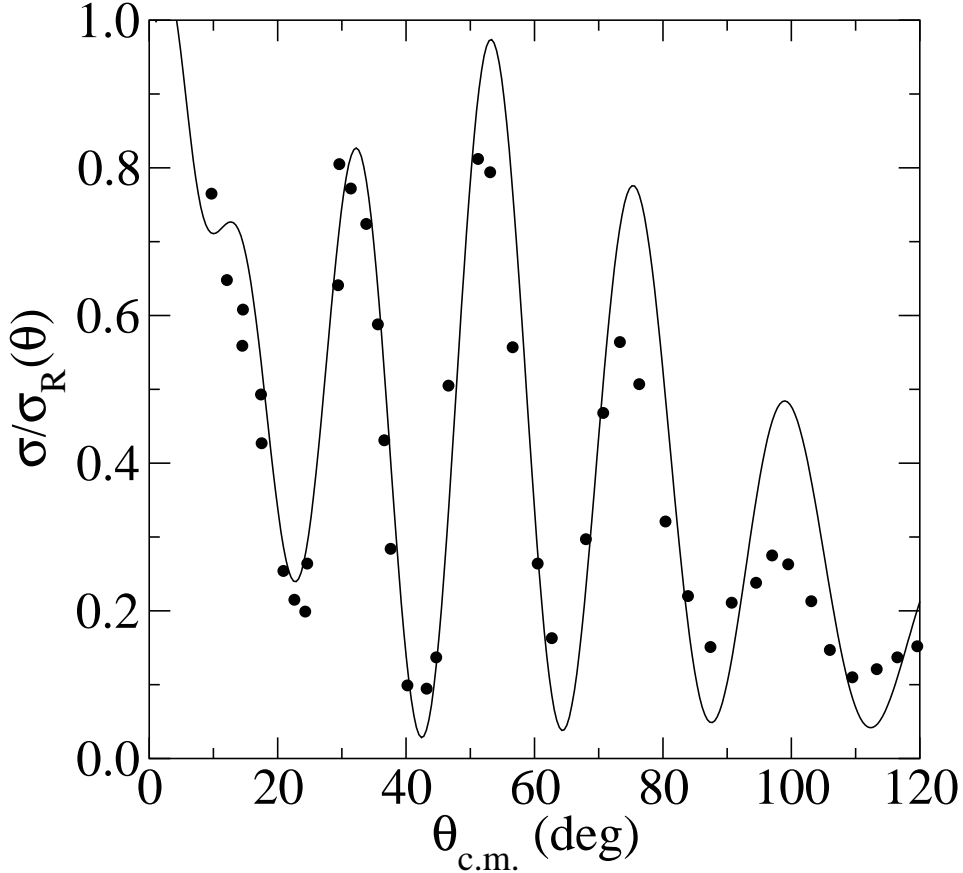}}
\caption{\label{Sn120-40-RR} 
The cross section as a ratio to Rutherford for the scattering of
39.8~MeV protons from $^{120}$Sn. The data~\cite{Fr67} are compared 
with the result of a calculation made using the $g$-folding optical 
potential as described in the text.}
\end{figure}

\begin{figure}
\scalebox{0.65}{\includegraphics*{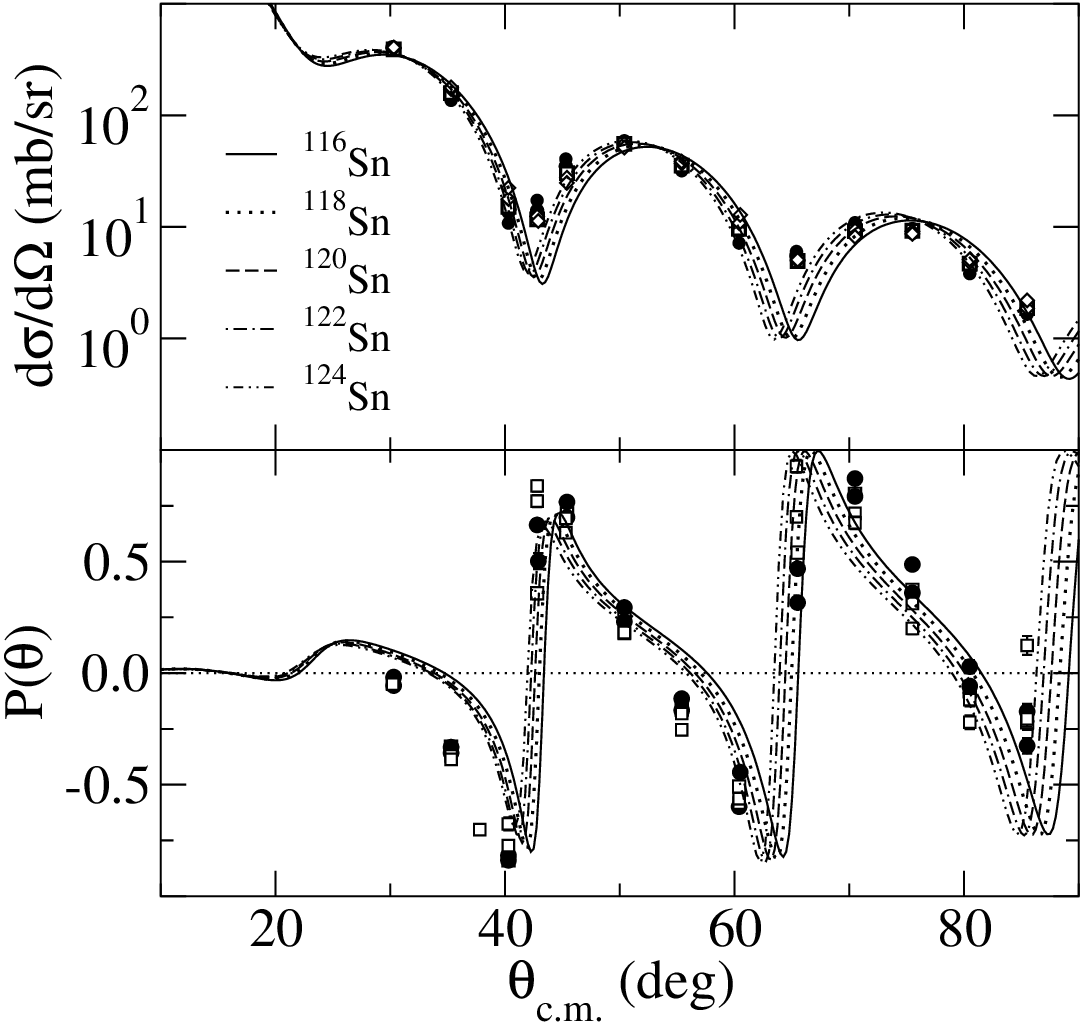}}
\caption{\label{Sn-cs+pol-40} The differential cross sections (top) and
polarizations (bottom) for the elastic scattering of 39.8~MeV protons
from $^{116,118,120,122,124}$Sn. The data~\cite{Bo68} are compared with
calculated results as found using the $g$-folding model described in the text.}
\end{figure}

\begin{figure}
\scalebox{0.65}{\includegraphics*{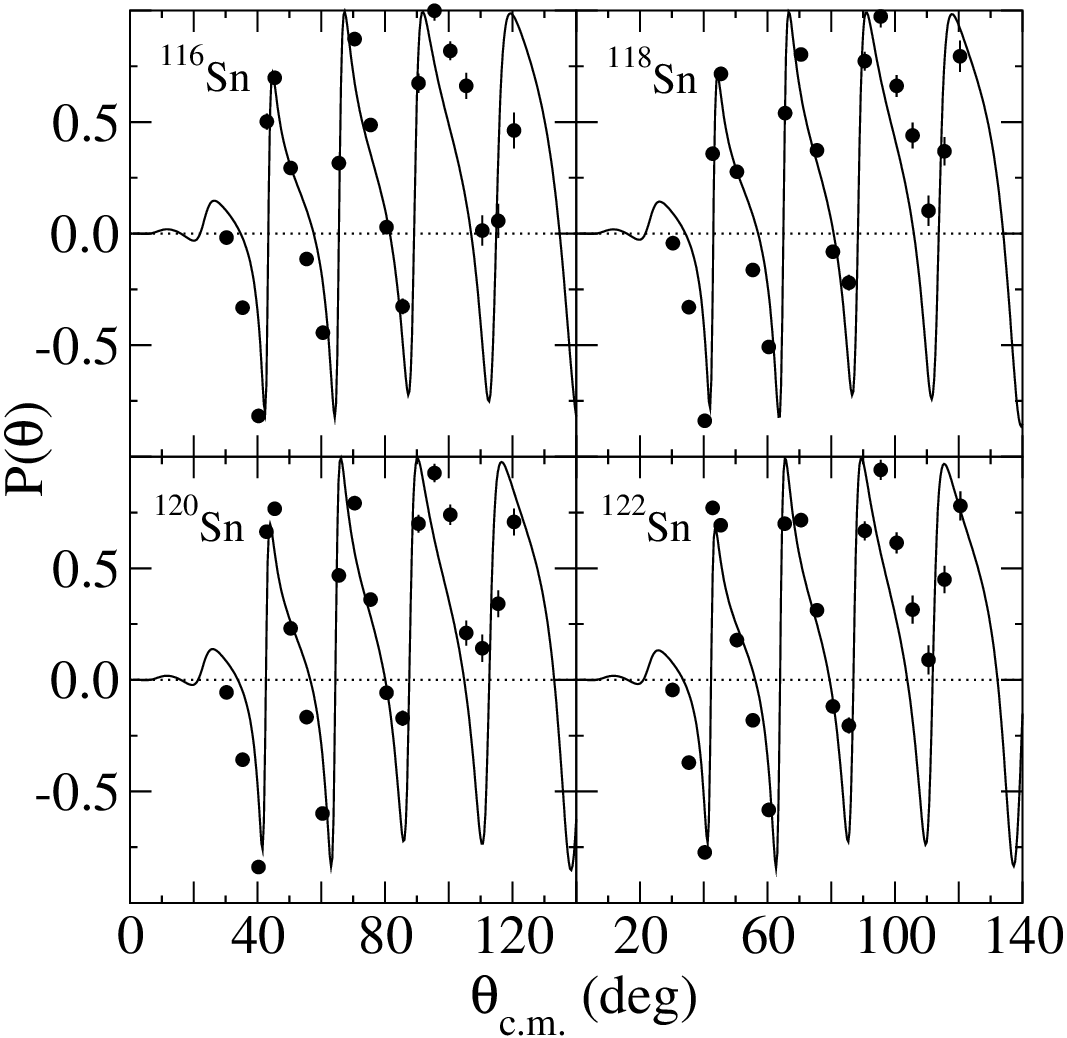}}
\caption{\label{Sn-39.8-pol}
Polarizations for the elastic scattering of 39.8~MeV protons
from $^{116,118,120,122}$Sn. The data~\cite{Bo68} are compared with
calculated results as found using the $g$-folding model described in the text.}
\end{figure}

\begin{figure}
\scalebox{0.65}{\includegraphics*{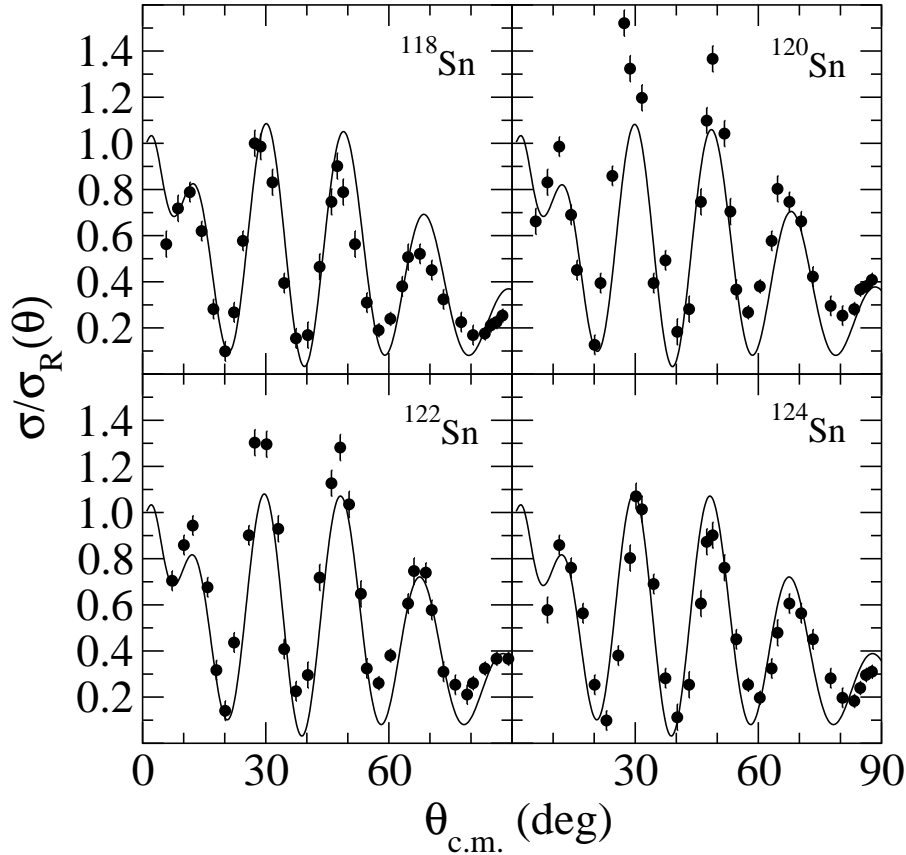}}
\caption{\label{Sn-50-RR} 
The cross section as a ratio to Rutherford for the scattering of
49.35~MeV protons from $^{118,120,122,124}$Sn. The data~\cite{Ma71} are compared 
with the results of calculations made using $g$-folding optical 
potentials as described in the text.}
\end{figure}

The first result we show is of the scattering of 39.8 MeV protons from
${}^{120}$Sn. In Fig.~\ref{Sn120-40-RR}, our calculated cross section
as a ratio to Rutherford scattering is compared with data~\cite{Fr67}.
Clearly the trend of the data is reproduced by the $g$-folding optical
potential calculation and some details are also well matched. Similar
quality of matching to data for the set of even mass isotopes,
${}^{116-124}$Sn, is revealed in Fig.~\ref{Sn-cs+pol-40}.
With all data collected in this one figure, and the spread of the calculated
values for the five isotopes giving a band of predictions, the trend of data 
is well reproduced by the  predicted values, at least to 90$^\circ$. 
The quality of fit to individual polarizations is presented in Fig.~\ref{Sn-39.8-pol}.
The maxima (positive and negative) values are in good agreement with
the characteristic shapes between the sharp rises in values (from negative to 
positive) being predicted. The variations between data and calculated values 
are not great and it seems within the realm of possibility that small changes to
the chosen structure model can improve the agreement.
Such is also the conclusion we may draw from the comparison of results
with data taken at 49.3 MeV.  Cross sections as ratios to Rutherford
are shown in Fig.~\ref{Sn-50-RR}. The data~\cite{Ma71} now display a
variation with target mass that is not readily mapped by our calculated
results. With ${}^{122,124}$Sn, the peak values of the ratios exceed 
our predictions by some 30 to 40\%.
However, the trend of data as well as a number of specific details
are well matched and we consider then that the results confirm
credibility of the SLy4 structure.

Finally we show results found for the elastic scattering of 65 MeV protons from
$^{118}$Sn and of 200 MeV protons from $^{120}$Sn.  At those energies many 
successful predictions have been made of cross sections for proton scattering 
from stable nuclei~\cite{Am00}.
Included in that set were predictions of scattering from two Sn isotopes
of the set considered herein. Those past Sn cross section calculations were
made using SP wave functions of HO type determined with an oscillator
length given by the  $A^{1/6}$ rule.  So values of  2.215 and 2.221~fm 
were used as the oscillator lengths for ${}^{118}$Sn and 
${}^{120}$Sn respectively.  In the following figures, we display cross 
sections found using those HO model densities by the dot-dashed curves.
Those results found using the SLy4 and SkP model densities are shown by 
the solid and dashed curves respectively.

\begin{figure}
\scalebox{0.65}{\includegraphics*{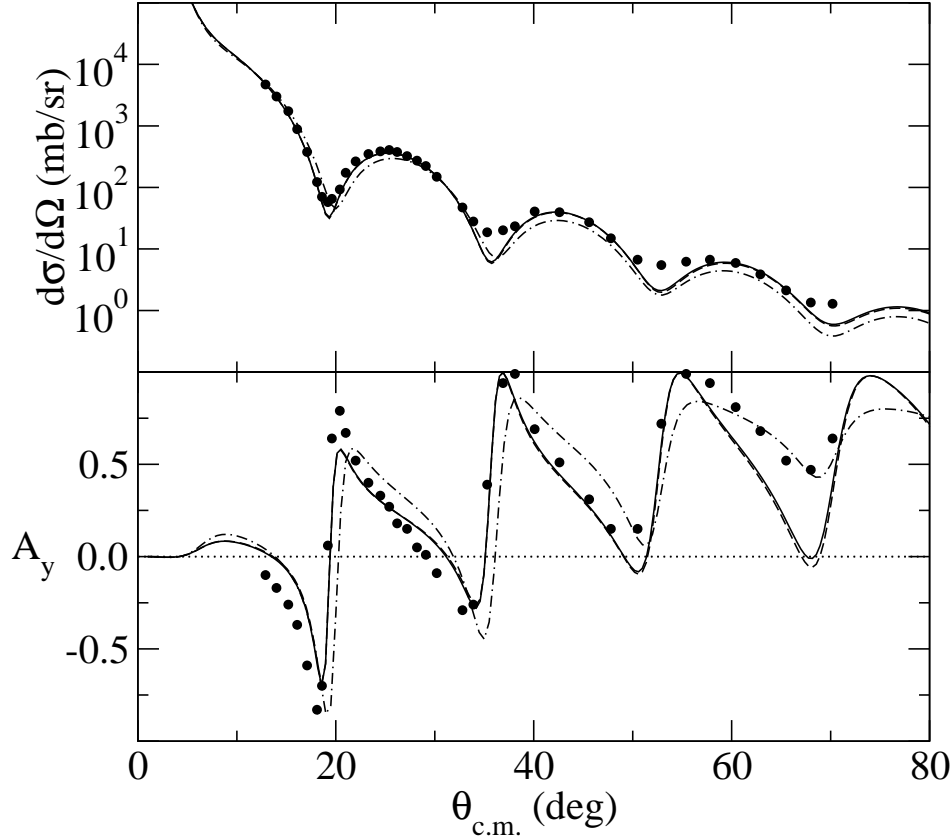}}
\caption{\label{sn118-65} The differential cross sections (top) and
analyzing powers (bottom) for the elastic scattering of 65~MeV protons
from $^{118}$Sn. The data \cite{Yo85} are compared to the results of
the calculations made using the SLy4 (solid line), SkP (dashed line)
and HO (dot-dashed line) models.}
\end{figure}

In Fig.~\ref{sn118-65}
the differential cross sections and analyzing powers for the elastic
scattering of 65~MeV protons from $^{118}$Sn~\cite{Yo85}, are compared 
with our predictions.
In the case of the differential cross section, there is little
difference ($\sim 1\%$) between those found using the SLy4 and SkP 
structures.  The calculated results are in good agreement with the
structure and magnitude of the data; the SLy4 and SkP model results
particularly so. The preference for the SLy4 (SkP) model also is
evident with the total reaction cross section.  They predict 1.48~b
while the HO calculation yields 1.41~b.  The measured
value~\cite{In99} is $1.535 \pm 0.047$~b.  But it is with the
analyzing power at 65~MeV that the SLy4 (SkP) densities give a
significantly better result in comparison to the HO model.  While
the HO result is a fair reproduction of the data structure, the SLy4
(SkP) model result gives not only the correct location of the maxima
and minima but also shows the overall trend, seen in the data, of
increasingly positive values with increasing angle. 
They also
depict best the marked asymmetric shape of each peak structure of the data.

The 200~MeV scattering results from $^{120}$Sn are also quite good,
although with the exception of the differential cross section produced
using the HO model the reproductions of data obtained with the
mean-field model densities are not as good as found at 65 MeV for
$^{118}$Sn.
\begin{figure}
\scalebox{0.65}{\includegraphics*{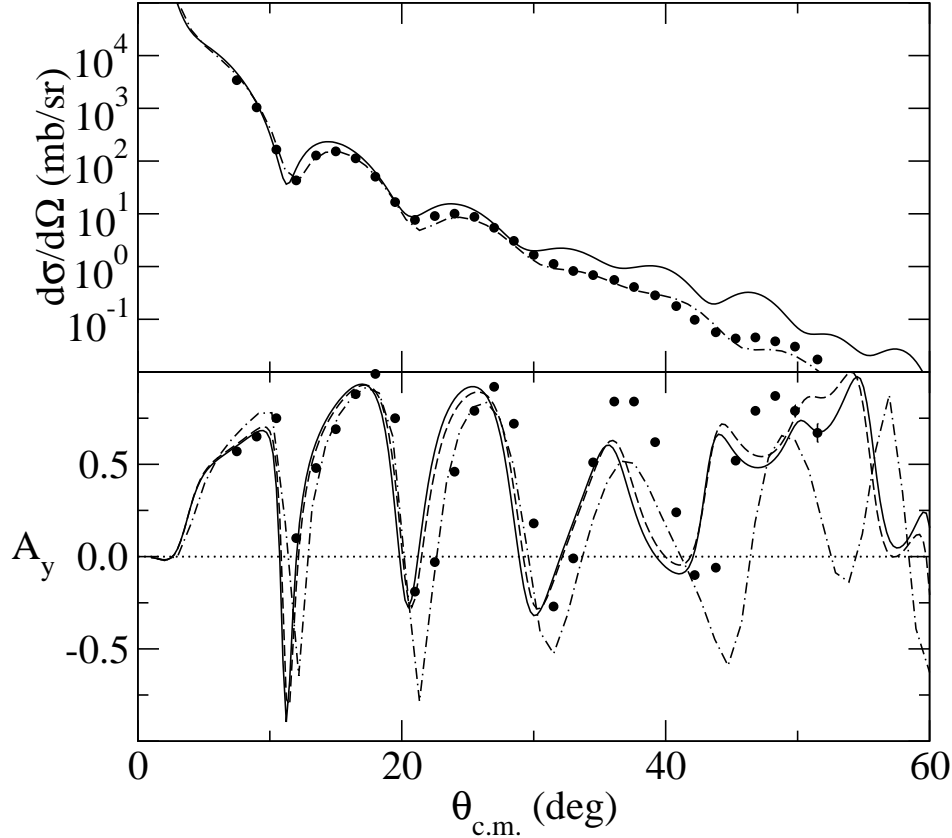}}
\caption{\label{sn120-200} The differential cross sections (top) and
analyzing powers (bottom) for the elastic scattering of 200~MeV
protons from $^{120}$Sn. The data \cite{Sa93} are compared to the
results of the calculations made as given in the text. The curves are
as for Fig.~\ref{sn118-65}. The SkP result is not given for
the cross section as it is indistinguishable from the SLy4 result.}
\end{figure}
The differential cross sections and analyzing powers are shown in
Fig.~\ref{sn120-200} from which it is clearly evident that the HO
model tracks the measured data~\cite{Sa93} well for all angles
shown. The SLy4 model result is not as good, though it agrees with the
forward angle ($< 30^{\circ}$) scattering data well enough. The SkP
model result is indistinguishable from the SLy4 one on this scale. The
actual differences in cross section are less than a percent at most
scattering angles. Both the SLy4 and SkP models thus 
overestimate the cross section and predict some structure not seen
in the data for larger scattering angles.  Such discrepancies have
been seen in other circumstances, notably in a comparative
study~\cite{Ka02} of model structures for $^{208}$Pb and in
identifying $^6$He and $^{11}$Li as nuclei with extended neutron
distributions (neutron halos) \cite{La01,St02}. Further,  RIA model
results for 200 and 295 MeV scattering~\cite{Te03} find
likewise predictions that are larger, and have more structure, than
the data for scattering angles greater than $\sim 15^\circ$.  It is also
of note that the 200 MeV data at scattering angles $> 25^\circ$ are smoothly 
decreasing and so quite unlike the equivalent data at lower energies as well as 
at 295 MeV~\cite{Te03}. However, it must be noted that the ``problematic" data
in this 200 MeV cross section has a magnitude $< 1$ mb/sr; cross section
magnitudes for which we have lessened confidence that processes other than
inherently considered within the $g$-folding method can be neglected; processes
such as may be associated with specific coupled channel for example.

As with the differential cross section results, our
predictions for the analyzing powers (shown in the bottom half of
Fig.~\ref{sn120-200}) are good but there is room for improvement. The
asymmetry seen with the 65~MeV data is less severe at this higher
energy but, of course, there are more peaks within the scattering
angle range shown. Both features are evident in the HO and SLy4
calculations without the exact angular structures in the data being
reproduced. It is notable that the SLy4 and SkP results do match the
observed peak magnitudes and valley depths very well and the two
densities yield slight but noticeable differences in the spin
measurable.

\section{Conclusions}
\label{conclude}

We have made predictions of the observables of the elastic scattering
of 200~MeV protons from the even-even isotopes of Sn from $^{100}$Sn
to $^{176}$Sn. The model we have used for the structure of the
isotopes is a Hartree-Fock-Bogoliubov model using the Skyrme
interaction for which two parameterizations have been used. The matter
densities obtained from those models show consistent trends. As mass
increases from $A = 100$ to 176, the neutron central density increases
and a neutron skin emerges. Also, as proton number is fixed, the
addition of neutrons not only increases the volume but also engenders
a dilution of the proton distribution.

The changes in the nucleon densities reflect in the predictions
made for the differential cross sections of 200~MeV proton elastic
scattering. As neutron number increases, the first three minima tend
to lower momentum transfers and the intervening first maximum becomes
more pronounced. That effect as the neutron skin becomes more
manifest is large enough to be distinguishable in experiment.

For the set of isotopes $^{116-124}$Sn cross sections, polarizations and
analyzing powers in proton elastic scattering have been measured at various 
energies. Most of that data compare well  with predictions from  $g$-folding 
optical potential calculations when those potentials are defined by folding with 
the SLy4 model of the structure of the isotopes.  However there is 
little difference between those results and ones obtained by using the 
SkP model.  But both of those models give results that are much better, usually, 
than those found by using a naive oscillator model of the structure of the isotopes.
The fact that there are marked differences 
between the HO and predictions signifies the value of using  proton elastic
scattering data in testing model specifications of the ground states
of nuclei.  The 200 MeV results for the scattering from ${}^{120}$Sn is exceptional.
Bearing in mind the effects observed in a recent study~\cite{Ka02}
of proton scattering from $^{208}$Pb, perhaps the neutron matter distribution through 
the nuclear surface as determined by either of the  calculations needs 
slight variation.

It is hoped that with the proposed new generation of radioactive beam
facilities that data for the elastic scattering of heavy neutron-rich
nuclei from hydrogen will be obtained. Despite the close similarity
of calculated results for the SLy4 and SkP model prescriptions,
the model we have used in obtaining
our predictions is sensitive to the major details in the neutron density and
serves as an important testing ground for the models being
developed to describe exotic nuclei and which inherently give
distinctive features such as the neutron rms value.

\begin{acknowledgments}
This work was supported by a grant from the Australian Research
Council and by the Polish
Committee for Scientific Research (KBN) under Contract
No.~5~P03B~014~21.  We also gratefully acknowledge the assistance of
Mr. Dirk van der Knijff of the Advanced Research Computing group,
Information Division, University of Melbourne for use of the
high performance computers of that group to find all of the results
displayed.
\end{acknowledgments}

\bibliography{Snpaper}

\end{document}